\documentstyle[preprint,aps,eqsecnum,epsfig]{revtex} 
\tighten 
\begin{document} 
\preprint{PITT-98-; CMU-HEP-98-02; DOE-ER/40682-142; LPTHE/98-09} 
\draft  
\title{\bf REAL-TIME RELAXATION   AND  KINETICS IN HOT
SCALAR QED: LANDAU DAMPING}  
\author{\bf Daniel Boyanovsky$^{(a)}$, H\'ector J. de Vega$^{(b)}$,
Richard Holman$^{(c)}$,  
S. Prem Kumar$^{(c)}$ and  Robert D. Pisarski$^{(d)}$}
\address
{(a) Department of Physics and Astronomy, University of 
Pittsburgh, Pittsburgh  PA. 15260, U.S.A \\ 
(b) LPTHE, Universit\'e Pierre et Marie Curie (Paris VI) et Denis Diderot 
(Paris VII), Tour 16, 1er. \'etage, 4, Place Jussieu, 75252 Paris, Cedex 05, 
France \\  
(c) Department of Physics, Carnegie-Mellon University, Pittsburgh, 
PA. 15213, U.S.A.\\  
(d) Department of Physics, Brookhaven National Laboratory, Upton,
NY 11973, U.S.A.}  
\date{February 1998}
\maketitle 
\begin{abstract} 
The real time evolution of non-equilibrium expectation values with soft length scales $\sim
k^{-1}>(eT)^{-1}$ is solved in hot scalar electrodynamics, with a view towards
understanding relaxational phenomena in the QGP and the electroweak plasma.
We find that the gauge invariant non-equilibrium expectation values relax via {\em power laws}
to asymptotic amplitudes that are determined by the quasiparticle  
poles. The long time relaxational dynamics and relevant time scales are
determined by the behaviour 
 of the retarded self-energy not at the small frequencies, but at the Landau
damping thresholds. This explains the presence of power laws and not
of exponential decay. In the process we rederive the HTL effective action using
{\em non-equilibrium} field theory. Furthermore we obtain the influence
functional, the Langevin equation and the fluctuation-dissipation theorem for
the soft modes, identifying the correlators that emerge in the classical
limit. We show that a Markovian approximation fails to describe the
dynamics {\em both} at short and long times. We also introduce a novel kinetic
approach that goes beyond the standard Boltzmann equation by incorporating
off-shell processes and find that the distribution function for soft
quasiparticles relaxes with a power law through Landau damping. 
We find an unusual dressing dynamics of bare particles and anomalous
(logarithmic) relaxation of hard quasiparticles.
\end{abstract} 
%\pacs{} 
\section{Introduction} 
%%%%%%%%%%%%%%%%%%%%%%%%%%%%%%%%%%%%%%%%%%%%
There is currently a great deal of interest in understanding non-perturbative
 real time 
 dynamics in gauge theories at high temperature, both within the realm of heavy
 ion collisions and the study of the quark gluon
 plasma \cite{qgp}-\cite{elze}, as well as the possibility for anomalous baryon
 number violation in the electroweak theory\cite{baryo,mcarnold}. In both
 situations the dynamics of soft gauge fields with typical length scales 
 $> (gT)^{-1}$ is non-perturbative. 

Their treatment requires a resummation  scheme 
 where one can consistently integrate out the hard scales associated with
 momenta $\approx T$ to obtain an effective theory for the soft
 scales. This is the program of resummation of hard thermal
 loops \cite{rob1}-\cite{blaizot1}.  
 Physically, the hard scale   
represents the typical energy of a particle in the plasma while
the soft scale   
is associated with collective excitations \cite{blaizot2}.

The recognition of the non-perturbative physics associated with soft degrees of
 freedom has led to an effort to describe the dynamics by
 implementing numerical simulations of 
{\em classical} gauge theories \cite{classi1}-\cite{classical} since soft
 degrees of freedom have very large occupation numbers and could in principle
 be treated classically \cite{classi4}. Effective classical
 descriptions for the 
 infrared bosonic modes have been obtained consistently in scalar field theory
 by integrating out the hard modes \cite{mullerclass}.  
 However, it was recognized that the dynamics of the soft modes in gauge
theories is sensitive to the hard modes \cite{bode,yaffe,yaff,son} and that the
Rayleigh-Jeans divergences associated with the hard modes provide non-trivial
contributions to the soft dynamics.   

For example, the one-loop correction to the  
gauge boson self-energy  contains a leading contribution from the hard momenta
of order $g^2T^2$ which gives the HTL (hard thermal loop) contribution,  and  
a subleading contribution from the soft scales. The $T^2$ 
dependence is a reflection of the UV quadratic divergence of the zero
temperature theory which is cutoff at momentum scales $> T$ by the
Bose-Einstein factor.  
Restoring the appropriate $\hbar's$ we see that this
contribution is of order $\hbar (g^2T^2/\hbar^2)$ where the first $\hbar$ is
associated with the loop and the 
denominator follows from the usual manner in which $\hbar$ enters with
temperature. This term is  therefore ${\cal O}(T^2/\hbar)$ and reveals the
usual Rayleigh-Jeans divergence.  
For hard external momenta  
($K_{ext}\sim T$) this one-loop correction to the propagator is obviously 
subleading and bare perturbation theory is valid. But when the external
momenta are soft  
($K_{ext}\sim gT$), clearly the one-loop correction is of the 
same order as the tree level term. This is in fact at the heart of the 
breakdown  
of the perturbative expansion. The problem is resolved by using HTL-resummed 
propagators and vertices for the soft external lines while hard scales may 
always be treated within the usual perturbation theory. This procedure is 
akin  
to obtaining a Wilsonian effective action for the soft modes by integrating 
out all the momenta above a certain soft scale which in this case is $gT$. For 
a detailed discussion of the relevant issues we refer the interested reader to 
the original works of Braaten and Pisarski \cite{rob1,bpisarski}. 
 
All of  the HTL contributions may be divided into two distinct categories: i) 
the  contributions from 
tadpole diagrams and ii) those from diagrams with discontinuities. While the 
tadpole contributions are independent of the external momenta, the
diagrams with  discontinuities  lead to momentum dependent terms  and it is
these that lead to the non-local effective HTL Lagrangian. The non-locality of
the HTL effective Lagrangian for 
the soft modes originates in the process of Landau damping which results in
discontinuities below the light
cone \cite{rob1}-\cite{rob2,htlabelian,htlnonabel,weldon1}. A very interesting
method to deal with the non-locality in the HTL effective action in numerical
simulations of classical gauge fields has been recently proposed
\cite{humuller} 
and is based on the particle method akin to that 
used in transport theory. This method has been used to study the
diffusion of Chern-Simons number in a lattice approach \cite{mooremuller}. A
local Hamiltonian approach that is intrinsically gauge invariant has also been
recently proposed \cite{iancu} and has the potential for numerical
implementation.  
A proposal to study the classical dynamics of soft gauge fields in terms of an
effective Langevin equation has been put forth in \cite{son}. However in
our 
view such a proposal does not seem to incorporate consistently the
non-Markovian nature of the  noise that is a result of the non-localities
associated with Landau damping.  

The focus of this article is precisely to study in detail the real-time
dynamics of the evolution of gauge fields, given an arbitrary 
field condensate in the initial state. This study will reveal that {\bf Landau
damping processes}   
dominate the most relevant  aspects of the dynamics and 
%originating from {\bf Landau damping processes} which, 
as argued above, determine the non-local aspects of the HTL
effective action. The main goal of this investigation is: i) to provide a
deeper 
understanding of the time scales associated with dissipative {\em off-shell}
processes, ii) a consistent microscopic description that can be used as a
yardstick to test lattice results on real-time correlation functions,
and iii) a detailed real-time description of relaxation and kinetics
of soft collective excitations in gauge theories.   
 
Landau damping \cite{blaizotqgp,rob1,bpisarski,rob2} 
occurs when a hard quasiparticle from the thermal bath 
(with momentum $\sim T$) scatters off a soft collective mode (momentum 
$\sim gT$), borrowing energy from (and damping) the soft
excitations in the process. Simple  
kinematics dictates that these processes can occur only {\em off shell} and
below 
the light-cone i.e. for  spacelike  
four-momentum. Furthermore, Landau damping  
gives a non-zero contribution only in the presence of a heat bath and when the 
external momentum is non-zero. Phrased differently, the Landau discontinuities 
are {\it purely thermal cuts} arising only at non-zero temperature and lead to 
damping of {\it spatially inhomogeneous} field configurations only. Whereas the
real-time dynamics of similar processes has been studied 
in a  scalar theory \cite{inhomodcc}, such a study is lacking for the case of
gauge fields. 
The gauge boson self-energy in the presence of 
these processes has been known for a long time
\cite{htlabelian,htlnonabel,weldon1} and  
can be computed  in 
the imaginary time formalism of finite temperature field theory in the HTL
limit\cite{lebellac}.    

\bigskip

%%%%%ADDITION 1

The main focus and goals of these article are:
\begin{itemize}

\item{ 

To compute explicitly the {\bf real time evolution} 
of inhomogeneous field configurations (non-equilibrium field expectation values) in the
ultrarelativistic plasma as an {\bf initial value problem}.  
We linearize the field equations of motion in the condensate amplitude.
In this weak field regime, the evolution equations for the condensate can be
solved in closed form through Laplace transform. The analytic
structure of the propagator in momentum space (s-plane) determines the
real time  behaviour of the solution. We obtain the propagator to
one-loop order in the HTL approximation.  
In this approximation the long-time behaviour of the condensate turns
out to be governed by the Landau   
discontinuities resulting in Landau damping processes.
Although the HTL corrections to the gauge boson self-energy are  
well-known, we believe that the  calculation of the {\em real
time} dependence of the damping of soft excitations is new. }

\item{
While an understanding of the real-time relaxation of non-equilibrium, 
inhomogeneous field configurations is of fundamental importance in the  physics
of relaxation in the QGP \cite{blaizotqgp},   
such a calculation also has  phenomenological implications 
for sphaleron induced B-violating processes. It has been 
recently pointed out \cite{yaffe,yaff,son} that standard estimates of the 
topological transition rate at finite temperature in the electroweak theory
ignore the effects of damping in the thermal bath and these authors have argued
that  Landau damping plays a very important role. Since the sphaleron is an   
inhomogeneous excitation associated with a soft length scale ($\sim1/g^2T$) 
Landau damping effects and the full HTL-resummed propagators must necessarily 
be taken into account when studying the sphaleron damping rates
\cite{yaffe,yaff,son}. 

%%%ADDITON 2 HERE

One of our main goals is to assess in detail  
 the real time non-equilibrium dynamics of soft excitations in the
 plasma with particular attention to a critical analysis of the
 long-standing 
belief that the small frequency region of the spectral function dominates
the long time relaxational dynamics. We find, to the contrary that the
Landau damping {\em thresholds} at $\omega = \pm k$ determine the long-
time dynamics and that the early time dynamics is sensitive to several
 moments of the total spectral density. This is an important point
 that bears on recent arguments 
that seek to clarify the damping effects on the sphaleron
 rate\cite{yaffe,yaff,son}. We analyze this novel result in detail both
 analytically and numerically, thus proving  that the
 long time behavior is dominated by the Landau damping {\em
 thresholds} and that the small frequency region gives rise to sub-leading
 corrections to 
the long-time dynamics in the leading order HTL approximation. }

\item{
 
In this article we concentrate on the case of scalar 
electrodynamics (SQED) since this theory has the same HTL structure (to lowest
order) as the non-abelian case \cite{lebellac,rebhan}. Most of our results can
therefore, be taken over to the
non-Abelian case  
with little or no changes at least in the lowest order HTL
approximation. Scalar 
electrodynamics has already been used as an  example to study the HTL
resummation for the infrared 
modes\cite{rebhan}, but the scope of this article is 
different in that we study explicitly the real time dynamics of the damping
processes.

After resummation of the one-loop HTL contributions, 
at long times the relaxation of either transverse or longitudinal
field expectation values is given by two contributions.  The first is
from the quasiparticle modes, and is standard --
an oscillatory function in time.
The second contribution arises from 
branch point singularities in the HTL self energies at
nonzero frequency.  These produce correlations in time
which are oscillatory times power law tails;
these power law tails are a new feature of HTL's.

%new paragraph%%%%
Long time power law tails in current-current correlators have been recently reported in reference\cite{yaff}. 
%end new paragraph

%!!!  A CHANGE HERE 

 }

\item{ 
Furthermore we obtain consistently the effective Langevin equation for soft
modes by integrating out the scalar fields and obtaining the influence
functional for the gauge invariant observables. This allows us to
extract the noise correlation function that displays all of the
non-localities associated with HTLs and Landau damping. By deriving the
relevant fluctuation-dissipation relation we identify the proper correlation
function that emerges in the classical limit.  
This analysis reveals  that the range of 
both the dissipation kernel as well as the noise correlation function are
determined 
by the {\em soft} scale. This results in the kernels being long-ranged,
typically falling off with a power of time and with no Markovian
limit. Again this result is deeply related to the non-localities of
the HTL effective action and prevents a local description of
relaxational dynamics associated with Landau damping.   

%!!!  A CHANGE HERE 

Our detailed analysis and the
consistent formulation of the influence functional starting from the
microscopic Lagrangian, unequivocally leads us to the conclusion that an 
effective stochastic Langevin description of gauge field relaxation to
leading order in the  HTL limit is {\em non-Markovian}. Because of the
long-range 
kernels associated with Landau damping, a Markovian limit cannot be
consistently extracted. This result points out the limited utility of
a Langevin equation for describing relaxation via Landau damping. } 

%%%%ADDITION 5

\item{ Having established the relaxation of inhomogenous gauge field
configurations via off-shell processes associated with Landau damping, we
ask how these processes contribute to the relaxation of the distribution
function of transverse degrees of freedom. Clearly the evolution of the
distribution function cannot be described to this order by a Boltzmann
equation, since this kinetic approach only includes on-shell processes. 
Thus we provide one of the novel results of this work:  we incorporate the non-equilibrium relaxation effects from Landau damping into a kinetic equation that describes the relaxation of the occupation number of transverse gauge
fields. This kinetic equation incorporates {\em off-shell} effects and
therefore constitutes an advance over the usual Boltzmann kinetic description
in terms of completed collisions.   
We argue that since Landau damping results in an exchange of
energy (and momentum) between the quasiparticles in the bath and the
out-of-equilibrium field configuration, this in turn will naturally lead to a
depletion of the particle number from the field configuration and must   
necessarily be included in    
any accurate kinetic description which aims to probe relaxational phenomena on 
the relevant time scales. A framework to study these transient,
off-shell relaxational phenomena as initial value problems is given in
refs.\cite{photop,boyakinetic,kinetics}. The analysis presented in 
this article implies a Dyson-like resummation that goes far beyond Boltzmann
kinetics. 

We compare the results from
these new kinetic equations to several different approximations to the
kinetics. In comparing the relaxation of dressed soft quasiparticles and
that of bare particles and hard quasiparticles, 
we find a remarkable dressing dynamics of the degrees of
freedom in the medium, and anomalous relaxation for the hard
quasiparticles. We also argue that a proper kinetic description 
must incorporate consistently the HTL effects and the quasiparticle nature of
the excitations.  }

\end{itemize} 

In Section II we introduce the 
model under study and focus on a gauge invariant description of the
dynamics. Section III is   
devoted to a derivation of the non-equilibrium equation of motion for the 
inhomogeneous condensate to one-loop order in the hard thermal 
approximation.  The real time relaxational
dynamics of the  
inhomogeneous configuration via Landau damping is investigated
in detail in Section IV. Contact is established with the
fluctuation-dissipation theorem and stochastic dynamics in Section V,
where we derive a Langevin description  
for the soft gauge invariant degrees of freedom in the thermal bath and
recognize the relevant correlation functions that emerge in a (semi) classical
stochastic description. We introduce a new kinetic description of transport
phenomena induced by the   
non-collisional Landau damping process in Section VI. In this section we study
the relaxation of the distribution function for soft quasiparticles, bare
particles and mention some interesting features of the 
relaxation for hard quasiparticles. Finally, we summarize our analysis,
discuss the modifications that will arise when higher order corrections 
leading to collisional lifetimes are included and
present our conclusions and possible future directions of study.

\section{Preliminaries}
As mentioned in the Introduction, our ultimate goal is to understand
relaxational processes associated with off-shell effects, such as Landau
damping, in a Non-Abelian gauge theory
\cite{blaizotqgp}. What we will do here is to treat the same problem in the
context of scalar quantum electrodynamics (SQED) model. To leading order, we expect
that this should be a good analogue of what happens in the Non-Abelian case; it
also has the advantage that it is simpler to deal with, and as we will see
below, it can be cast from the outset in terms of {\em gauge invariant}
variables. This will eliminate any ambiguities associated with the usual
problem of gauge dependence of off-shell quantities.

We will start with some inhomogeneous field configuration which is excited in the
SQED plasma at $t=0$. What we want to do is to follow the time development of
this configuration as it interacts with the hard modes in the plasma, and in
particular, we want to know whether the relaxation is of the usually assumed
exponential sort or something different. 

Let us first reformulate SQED in terms of gauge invariant variables. 
The SQED Lagrangian is given by   
\begin{eqnarray} 
&&{\cal{L}}=D_\mu\Phi^\dagger\,D^\mu\Phi-m^2|\Phi|^2-\frac{1}{4}F_{\mu\nu} 
F^{\mu\nu}\; ,\nonumber \\ 
&&D_\mu\Phi=(\partial_\mu-ieA_\mu)\Phi\; . 
\end{eqnarray} 
 
A description of the dynamics in terms of gauge invariant observables begins
with the identification of the constraints associated with gauge
invariance. The Abelian gauge theory has two first class constraints, namely
Gauss's law and vanishing canonical momentum for $A_0$. What we will do is
project the theory directly onto the physical Hilbert space, defined as usual
as the set of states annihilated by the constraints. The procedure is simple.
First, we obtain gauge invariant observables that commute with the first class
constraints and write the Hamiltonian in terms of these. All of the matrix
elements between gauge invariant states (annihilated by first class
constraints) are the same as those that would be obtained by fixing Coulomb
gauge $\vec{\nabla}\cdot\vec{A}=0$. The Hamiltonian, when defined in the
physical subspace, can be written solely in terms of transverse components and
includes the instantaneous Coulomb interaction as would be obtained in Coulomb
gauge.  This instantaneous Coulomb interaction can then be traded for a {\em
gauge invariant} Lagrange multiplier field $A_0(\vec x,t)$ (a non-propagating
field whose canonical momentum is absent from the Hamiltonian) linearly coupled
to the charge density $\rho(\vec{x}, t)$ and obeying the algebraic equation of
motion $\nabla^2 A_0(\vec x,t)=\rho(\vec x,t)$. Alternatively, one can use a
phase-space path integral representation of the generating functionals, trade
the Coulomb interaction with a Lagrange multiplier linearly coupled to the
charge density and perform the path integral over the canonical momenta as
usual. Both methods lead to the following Lagrangian density:
\begin{eqnarray}  
{\cal L}=&&\partial_\mu\Phi^\dagger\,\partial^\mu\Phi 
+\frac{1}{2}\partial_\mu \vec{A}_T\cdot\partial^\mu\vec{A}_T 
-e\vec{A}_T\cdot\vec{j}_T 
-e^2\vec{A}_T\cdot\vec{A}_T\; \Phi^\dagger\Phi +\nonumber \\ 
&&+\frac{1}{2}\left(\nabla A_0 \right)^2+ {e^2}A^2_0 \; \Phi^{\dagger}\Phi-i 
eA_0\;\left(\Phi
\dot{\Phi}^{\dagger}-{\Phi}^{\dagger}\dot{\Phi}\right) \; , \nonumber \\
%\label{lagrainva}\\ 
\vec{j}_T=&&i(\Phi^\dagger\vec{\nabla}_T\Phi-\vec{\nabla}_T\Phi^\dagger~\Phi)\;.  
\label{current} 
\end{eqnarray} 
\noindent where $A_T$ is the transverse component of the gauge field. 

In order to provide an initial value problem for studying the relaxational
dynamics of charge density fluctuations we introduce an external source $ {\cal
J}_L(\vec x,t) $ linearly coupled to $A_0$ and study the linear response to
this perturbation. Furthermore, it is convenient to introduce external sources
coupled to the transverse gauge fields to study the linear response of {\em
transverse} gauge field configurations.  These external fields could in
principle play the role of a semiclassical configuration coupled to small
perturbations in a linearized approximation. Therefore we include external
source terms in the Lagrangian density, 
\begin{equation} \label{fuente}
{\cal L} \rightarrow {\cal L}-{\cal J}_L(\vec x,t) A_0(\vec
x,t)-\vec{{\cal J}}_T(\vec x,t)\cdot \vec{A}_T(\vec x,t) \; . 
\end{equation}  

The relaxational dynamics of our initial inhomogeneous configurations is
clearly an out of equilibrium process, and needs to be treated by an
appropriate formalism\cite{ctp,disip,tadpole}.

In the Schr\"odinger picture the dynamics is completely described by a
time-evolved density matrix $\hat{\rho}$ that obeys the quantum Liouville
equation:
\begin{equation} 
i\hbar\frac{\partial\hat{\rho}}{\partial t}=[\hat{\rho},H]\;, 
\end{equation}  
where $H$ is the Hamiltonian  of the system. The expectation value of any 
operator ${\cal O}$ is given by
\begin{equation} 
\langle{\cal O}\rangle=\text{Tr}[\hat{\rho}(t){\cal O}]\;. 
\end{equation} 
For thermal initial conditions with an initial temperature given by $1/\beta$,
the density matrix at $t=0$ is $\rho_i=e^{-\beta H_i}$ and the above
expectation value can be rewritten easily as a functional integral defined on a
complex-time contour. Notice that $ H_i $ {\em is not} the Hamiltonian of the
system for $t>0$. The system thus evolves out of equilibrium.  The contour has
two branches running forward and backward in time and a third leg along the
imaginary axis stretching to $t=-i\beta$. This is the standard
Schwinger-Keldysh closed time path formulation of non-equilibrium field theory
(see ref. \cite{ctp,disip,tadpole} for details). Fields defined on the forward
and backward time contours are accompanied with $(+)$ and $(-)$ superscripts
respectively and are to be treated independently. The expectation value of any
string of field operators may be obtained by introducing independent sources on
the forward and backward time contours and taking functional derivatives of the
generating functional with respect to these sources. The imaginary time leg of
the complex time contour does not contribute to the dynamics.  Since the path
integral represents a trace, the initial and final states must be identified
and therefore all the local bosonic fields ${{\cal O}(\vec{x},t)}$ satisfy the
Kubo-Martin-Schwinger (KMS) periodicity condition
\begin{eqnarray} 
{\cal O}^{(+)}(\vec x, t_0)={\cal O}^{(\beta)}(\vec x,t_0-i\beta). 
\end{eqnarray}    
The non-equilibrium SQED Lagrangian is given by 
\begin{equation} 
{\cal L}_{noneq}= {\cal L}\left[\vec A_T^+,\Phi^+,\Phi^{\dagger +},A_0^+\right]
- {\cal L}\left[\vec A_T^-,\Phi^-,\Phi^{\dagger -},A_0^-\right]\;. 
\end{equation} 
 
Perturbative calculations are carried out with the following non-equilibrium
Green's functions:
 
\begin{itemize}
\item{ Scalar Propagators:

$${\langle}{\Phi}^{(a)\dagger}(\vec{x},t){\Phi}^{(b)}(\vec{x}, 
t^{\prime}){\rangle}=-i\int {d^3k\over{(2\pi)^3}}\; G_k^{ab}(t,t^\prime) \;
e^{-i\vec{k}\cdot(\vec{x}-\vec{x^\prime})}\;, 
$$
where $(a,b)\;\in\{+,-\}$. 
\begin{eqnarray} 
&&G_k^{++}(t,t^\prime)=G_k^{>}(t,t^{\prime})\Theta(t-t^{\prime}) 
+G_k^{<}(t,t^{\prime})\Theta(t^{\prime}-t)\; , \label{gplpl} 
\\ 
&&G_k^{--}(t,t^\prime)= G_k^{>}(t,t^{\prime})\Theta(t^{\prime}-t)+ 
G_k^{<}(t,t^{\prime})\Theta(t-t^{\prime})\;, \nonumber
\\ 
&&G_k^{\pm\mp}(t,t^\prime)=-G_k^{<(>)}(t,t^{\prime})\;, 
\label{gplmin}\\ 
&&G_k^{>}(t,t^{\prime})=\frac{i}{2\omega_k}\left[ 
(1+n_k)\;e^{-i\omega_k(t-t^\prime)} 
+n_k\;e^{i\omega_k(t-t^\prime)}\right],\label{greater} 
\\ 
&&G_k^{<}(t,t^{\prime})=\frac{i}{2\omega_k}\left[ 
n_k\;e^{-i\omega_k(t-t^\prime)} 
+(1+n_k)\;e^{i\omega_k(t-t^\prime)}\right]\;,\label{lesser} 
\\ 
&&\omega_k=\sqrt{\vec{k}^2+m^2}\quad;\;\;\;\;\;\;n_k= 
\frac{1}{e^{\beta\omega_k}-1}\;.  \nonumber
\end{eqnarray}} 
\item{Photon Propagators:

$${\langle}{A}^{(a)}_{Ti}(\vec{x},t){A}^{(b)}_{Tj}(\vec{x}, 
t^{\prime}){\rangle}=-i\int {d^3k\over{(2\pi)^3}}\;{\cal G}_{ij}^{ab} 
(k;t,t^\prime)\;e^{-i\vec{k}\cdot(\vec{x}-\vec{x^\prime})}\;, 
$$
\begin{eqnarray} 
&&{\cal G}_{ij}^{++}(k;t,t^\prime)={\cal P}_{ij}(\vec{k}) \;
\left[{\cal G}_k^{>}(t,t^{\prime})\Theta(t-t^{\prime}) 
+{\cal G}_k^{<}(t,t^{\prime})\Theta(t^{\prime}-t) \right]\;, \nonumber
\\ 
&&{\cal G}_{ij}^{--}(k;t,t^\prime)= {\cal P}_{ij}(\vec{k}) \;
\left[{\cal G}_k^{>}(t,t^{\prime})\Theta(t^{\prime}-t) 
+{\cal G}_k^{<}(t,t^{\prime})\Theta(t-t^{\prime}) \right]\;, \nonumber
\\ 
&&{\cal G}_{ij}^{\pm\mp}(k;t,t^\prime)=-{\cal P}_{ij}(\vec{k}) \;
{\cal G}_k^{<(>)}(t,t^{\prime})\;, \nonumber
\\ 
&&{\cal G}_k^{>}(t,t^{\prime})=\frac{i}{2k}\left[ (1+N_k)e^{-ik(t-t^\prime)} 
+N_ke^{ik(t-t^\prime)}\right]\;,\label{phot>} 
\\ 
&&{\cal G}_k^{<}(t,t^{\prime})=\frac{i}{2k}\left[ 
N_k\;e^{-ik(t-t^\prime)} 
+(1+N_k)\;e^{ik(t-t^\prime)}\right],\label{phot<}\\ 
&&N_k=\frac{1}{e^{\beta k}-1}\;. \nonumber
\end{eqnarray}}
\end{itemize}
Here ${\cal P}_{ij}(\vec{k})$ is the transverse projection operator: 
\begin{equation} 
{\cal P}_{ij}(\vec{k})=\delta_{ij}-\frac{k_ik_j}{k^2}\;.\label{projector} 
\end{equation}  
 
With these tools we are ready to begin our analysis of non-equilibrium SQED.

\section{Linear Relaxation}

We now introduce the inhomogeneous non-equilibrium expectation values 
$\vec{\cal A}_{T}(\vec{x},t) \; ; \; {\cal A}_0(\vec x,t)$ which are excited at time $t=0$ in the plasma. The
non-equilibrium expectation values of the transverse components represent electric and magnetic
fields, whereas the expectation of the Lagrange multiplier field
$A_0$ corresponds to an initial charge density in the system.

The dynamics of these non-equilibrium expectation value will be analyzed by treating $\vec{\cal
A}_{T}(\vec{x},t); \, {\cal A}_0(\vec x,t)$ as background fields, i.e. the
expectation values of the corresponding fields in the non-equilibrium density
matrix, and expanding the Lagrangian about this configuration. Therefore we
split the full quantum fields into c-number expectation values (which are the
non-equilibrium expectation value) and quantum fluctuations about these expectation values:
\begin{eqnarray} 
&&\vec{A}_T^{(\pm)}(\vec{x},t)=\vec{{\cal A}}_T(\vec{x},t) 
+\delta\vec{A}_T^{(\pm)}(\vec{x},t) \; ; \;  
{A}_0^{(\pm)}(\vec{x},t)={{\cal A}}_0(\vec{x},t) 
+\delta{A}_0^{(\pm)}(\vec{x},t) 
\label{condensate1}\\ 
&&\vec{{\cal A}}_T(\vec{x},t)=\langle\vec{A}_T^{(\pm)}(\vec{x},t)\rangle 
\quad ; \quad {{\cal A}}_0(\vec{x},t)=\langle {A}_0^{(\pm)}(\vec{x},t)\rangle 
\label{condensate2} 
\end{eqnarray} 
where the expectation values of the field operators are taken in the
time-evolved density matrix.

The equations of motion for the background field can be obtained to any order
in the perturbative expansion by imposing the requirement that the expectation
value of the quantum fluctuations in the time evolved density matrix vanishes
identically.  This is referred to as the tadpole equation \cite{tadpole} which
follows from Eq.(\ref{condensate1}) and Eq.(\ref{condensate2}):
\begin{equation} 
\langle \delta \vec{A}_T^{(\pm)}\rangle=0 \quad ; \quad \langle \delta
{A}_0^{(\pm)}\rangle=  0 \label{tadpoleq}\;.  
\end{equation} 
The equations obtained via this procedure are the equations of motion obtained
by variations of the non-equilibrium effective action. The perturbative
expansion needed to compute the relevant expectation values is obtained by
treating {\it all} the {\it linear} terms \cite{tadpole} in the fluctuations as
interactions along with the usual interaction vertices. 

Although in principle the tadpole method could be used to study arbitrary
background configurations including non-perturbative ones (for e.g. sphalerons
in the non-Abelian case) we restrict our discussions to the small amplitude
regime in close analogy to the work of ref.\cite{inhomodcc} for scalar field
theory. In other words, the effective action equation of motion will be studied
in the linear approximation for the condensate amplitude so that $ {\cal
O}\left({\cal A}_{Ti}^2\right) $ and $ {\cal O}\left( {\cal A}_{0}^2\right) $
and higher orders will be neglected. In addition, the evolution kernel will be
approximated to the {\it one-loop} order only.

Defining the Fourier components of the electromagnetic condensate as 
 
\begin{eqnarray} 
&&{\cal A}_{Ti}(\vec{k},t)=\int d^3x\;e^{i\vec{k}\cdot\vec{x}} \;
{\cal A}_{Ti}(\vec{x},t),\\ 
&&{\cal A}_{0}(\vec{k},t)=\int d^3x\;e^{i\vec{k}\cdot\vec{x}} \;
{\cal A}_{0}(\vec{x},t) 
\end{eqnarray} 
we obtain the following 
equation of motion to one-loop order for the transverse part: 
 
\begin{eqnarray} 
&&\frac{d^2}{dt^2}{\cal A}_{Ti}(\vec{k},t)+k^2{\cal 
A}_{Ti}(\vec{k},t)+2e^2\langle\Phi^\dagger\Phi\rangle
\;{\cal A}_{Ti}(\vec{k},t)+  \nonumber \\ \nonumber 
&&-2e^2\int_0^td\tau\int\frac{d^3p} {(2\pi)^3\omega_p\omega_{p+k}}\;
p_{Ti}\;p_{Tj}\; [(1+n_p+n_{p+k})\sin\{(\omega_{k+p}+  
\omega_p)(t-\tau)\}\\
%&&\hspace{2 in}
&&+(n_p-n_{k+p})\sin\{(\omega_{k+p}-\omega_p)(t-\tau)\}] \;
{\cal A}_{Tj}(\vec{k},\tau)={\cal J}_{Ti}(\vec k,t) \; . \label{transeqmotion1}
\end{eqnarray} 
Here $p_{Ti}$ refers to the component of the spatial momentum $\vec{p}$ which
is transverse to the wave-vector $\vec{k}$ and $ {\cal A}_{Tj}(\vec{k},\tau) $
stands for the Fourier transform of the external source $ \vec{A}_{Tj}(\vec
x,t) $ in Eq.(\ref{fuente}).

The tadpole term which appears due to the 4-point `seagull' vertex can be
evaluated easily in the limit where $T>>m$ and yields the following hard
thermal loop contribution
 
\begin{eqnarray} 
2e^2\langle\Phi^\dagger\Phi\rangle=2e^2\int\frac{1+2n_k}{2\omega_k}\frac{d^3k} 
{(2\pi)^3}\simeq\frac{e^2T^2}{6}\;.\label{tadpolemass} 
\end{eqnarray} 
 
We remark that in evaluating the tadpole diagram above and in all subsequent
calculations it is always {\it assumed} that the zero temperature divergences
have already been absorbed in the proper mass and wavefunction
renormalizations.
 
The longitudinal part obeys a similar  equation: 

\begin{eqnarray} 
&& k^2{\cal A}_{0}(\vec{k},t)+e^2\int_0^td\tau\int\frac{d^3p} 
{(2\pi)^3}\left[\left(\frac{\omega_{k+p}}{\omega_p}-1\right) 
(1+n_p+n_{p+k})\sin\{(\omega_{k+p}+\omega_p)(t-\tau)\}
\right. \nonumber \\ 
&&\left.+\left(\frac{\omega_{k+p}}{\omega_p}+1\right) 
(n_p-n_{k+p})\sin\{(\omega_{k+p}-\omega_p)(t-\tau)\}\right] 
{\cal A}_{0}(\vec{k},\tau)={\cal J}_L(\vec k,t)\;.\label{longeqmotion1} 
\end{eqnarray} 
It should be noted that the equation of motion for the longitudinal component
has no time derivatives, indicating the non-dynamical nature of the field. The
source term for the longitudinal component is interpreted as an external
disturbance that induces a charge density fluctuation in the SQED plasma.  The
response to a general disturbance can be obtained by convolution from the
result of linear response to the impulsive perturbation.
 
We also note that unlike the transverse case, there is no contribution from the
tadpole term to the effective action equations of motion.  This fact will be
important in understanding the origin of a non-zero Debye mass.
 
The nonlocal terms in Eqs.(\ref{transeqmotion1}) and (\ref{longeqmotion1}) are
the one-loop self-energies for the transverse and longitudinal components
respectively, resulting from the photon-Higgs trilinear coupling. The first
piece in the nonlocal terms proportional to $ 1+n_p+n_{k+p} $ is the difference
of the following creation and annihilation processes in the medium
\cite{inmedium} : $ \gamma\rightarrow\bar{\Phi}\;\Phi $ with Bose factor $(
1+n_p)\;(1+n_{p+k})$ and $\bar{\Phi}\; \Phi\rightarrow\gamma$ with a
statistical factor $n_p\;n_{p+k}$.  The piece proportional to $(n_p-n_{k+p})$
is the Landau damping contribution \cite{rob2,inmedium}. When $k=0$ this
contribution vanishes indicating that it affects inhomogeneous excitations
only. Furthermore, it has no zero temperature counterpart since the Bose factor
$ n_k-n_{k+p} $ vanishes identically at $T=0$. This term arises from the
difference of the processes $\gamma\; \Phi\rightarrow\Phi$ with statistical
factor $(1+n_{k+p})\;n_p$ and $\Phi\rightarrow
\gamma\;\Phi$ with the factor $(1+n_p)\;n_{k+p}$.  
Although it does not give rise to an imaginary part for the {\it 
on-shell} self-energy, it will nevertheless have an effect on the physical
processes associated with the relaxation of the inhomogeneous condensate as
described in detail below.   
 
We can solve Eqs.(\ref{transeqmotion1}, \ref{longeqmotion1}) via the Laplace
transform. Introducing the Laplace transformed fields  

\begin{eqnarray} 
&&\tilde{\cal A}_{Ti}(\vec{k},s)=\int_0^\infty dt\; e^{-st} 
{\cal A}_{Ti}(\vec{k},t)\; ,\\ 
&&\tilde{\cal A}_{0}(\vec{k},s)=\int_0^\infty dt\; e^{-st} 
{\cal A}_{0}(\vec{k},t)\;, 
\end{eqnarray} 
and performing the transform on the above equations of motion we get following
the same methods as in ref.\cite{inhomodcc},

\begin{eqnarray}
\nonumber\\
&&(s^2+k^2+e^2T^2/6)\; \tilde{{\cal A}}_{Ti}(\vec{k},s)+\label{transeqmotion} 
\\\nonumber 
&&-2e^2\int \frac{d^3p}{(2\pi)^3 
\omega_p\omega_{k+p}}\left[(1+n_p+n_{p+k})\frac{\omega_{k+p}+\omega_p} 
{s^2+(\omega_{k+p}+\omega_p)^2}+\right. \\\nonumber  
&&\left. 
+(n_p-n_{k+p})\frac{\omega_{k+p}-\omega_p} 
{s^2+(\omega_{k+p}-\omega_p)^2}\right] 
p_{Ti}\;p_{Tj}\;\tilde{{\cal A}}_{Tj}(\vec{k},s)  
=s{\cal A}_{Ti}(\vec{k},0)+\dot{{\cal A}}_{Ti}(\vec{k},0) + \tilde{\cal
J}_{Ti}(\vec k,s)  \\ \nonumber 
\end{eqnarray} 
for the transverse part, and  
\begin{eqnarray} 
\nonumber\\ 
&&\left\{k^2 
+e^2\int \frac{d^3p}{(2\pi)^3}\left[\left(\frac{\omega_{k+p}} 
{\omega_p}-1\right) (1+n_p+n_{p+k})\frac{\omega_{k+p}+\omega_p} 
{s^2+(\omega_{k+p}+\omega_p)^2}+\right.\label{longeqmotion}\right.
\\\nonumber  
&&\left. \left.
+\left(\frac{\omega_{k+p}}{\omega_p}+1\right) 
(n_p-n_{k+p})\frac{\omega_{k+p}-\omega_p}  
{s^2+(\omega_{k+p}-\omega_p)^2}\right] \right\}
\tilde{{\cal A}}_{0}(\vec{k},s)  =\tilde{\cal J}_L(\vec k,s) \; . 
\\\nonumber 
\end{eqnarray} 
for the longitudinal component. The Laplace transform variable $s$ plays the
role of an (imaginary) time component of the photon four-momentum.

\section{Real Time Landau Damping} 
 
With the equations of motion for the non-equilibrium expectation value in hand, we turn to the
analysis of its time evolution, paying particular
attention to those parts of the spectral density which contribute to the long
time dynamics.

%In this section we analyze the time evolution of the gauge condensate  with
%particular attention to the contribution to the long time dynamics from
%different parts of the spectral density. 
 
%Having derived relevant results for Abelian theories at finite temperature
%using the real-time non-equilibrium framework, we now turn to the main issues

%of this work, namely to the detailed study of the dynamics of relaxation via
%Landau damping.

We do this by solving Eqs.(\ref{transeqmotion}, \ref{longeqmotion}) via the
inverse Laplace transform. Recall that this requires an integration along a
contour parallel to the imaginary $s$ axis placed in such a way so as to have
all the singularities of the integrand to the left of the contour. To do this,
we will need a clear understanding of the analytic structure of the photon
self-energy in the $s$-plane. We will explicitly outline the details of the
calculation for the transverse part only, since those for the longitudinal part
are similar.
 
\subsection{The Transverse Part:} 

It is illuminating to write the
equation of motion (\ref{transeqmotion}) in terms of a
spectral density function $ \rho_{ij}  (\omega^\prime,\vec{k}) $
\cite{inhomodcc},   
\begin{eqnarray} 
&&(s^2+k^2+e^2T^2/6)\tilde{{\cal A}}_{Ti}(\vec{k},s)+\int_{-\infty}^{+\infty} 
d\omega^\prime\frac{\omega^\prime} 
{s^2+\omega^{\prime 2}} \; 
\rho_{ij}(\omega^\prime,\vec {k}) \;
\tilde{{\cal A}}_{Tj}(\vec{k},s)\\\nonumber  &=& 
s {\cal A}_{Ti}(\vec{k},0) + \dot{{\cal A}}_{Ti}(\vec{k},0) +
\tilde{\cal J}_{Ti}(\vec k,s)  \\ \nonumber 
\end{eqnarray} 
where 
\begin{eqnarray} 
\rho_{ij}(\omega^\prime,\vec{k})&=& 
-2e^2\int \frac{d^3p}{(2\pi)^3\omega_p\omega_{k+p}} 
p_{Ti}\;p_{Tj}\times\\\nonumber 
&&\times\left[(1+n_p+n_{p+k})\delta(\omega^\prime-\omega_{k+p}-\omega_p) 
+(n_p-n_{k+p})\delta(\omega_{k+p}-\omega_p-\omega^\prime)\right]. 
\end{eqnarray} 
The one-loop transverse self-energy in the $s$-plane is given by:   
\begin{equation} 
{\cal P}_{ij}(\vec{k})\Sigma^{t}(s)=\frac{e^2T^2}{6}\;{\cal P}_{ij}(\vec{k}) 
+\int_{-\infty}^{\infty}d\omega^\prime 
\frac{\omega^\prime}{s^2+\omega^{\prime 2}}\;\rho_{ij} 
(\omega^\prime,\vec{k},s),\label{selfenergy} 
\end{equation} 
where we have recognized the  fact that $\rho_{ij}\propto{\cal
P}_{ij}(\vec k)$.    
 
Using ${\cal P}_{ij}(\vec k){\cal A}_{Tj}(\vec k,s)={\cal A}_{Ti}(\vec
k,s)$ the  Laplace transform of the transverse part of the condensate
is given by:  
\begin{equation} 
\tilde{\cal A}_{Ti}(\vec{k},s)= 
\frac{s \, {\cal A}_{Ti}(\vec{k},0)+\dot{\cal A}_{Ti}(\vec{k},0)
+\tilde{\cal J}_{Ti}(\vec k,s)}  
{s^2+k^2+\Sigma^t(s)} 
\end{equation} 
 The real time dependence of the inhomogeneous background is given by
the inverse Laplace  
transform of the above expression which is in fact the retarded propagator 
defined in the $s$-plane. The inverse transform is calculated by
performing the following integral along the Bromwich contour.
\begin{eqnarray} 
{\cal A}_{Ti}(\vec k, t)=&&\int_{c-i\infty}^{c+i\infty}
\frac{ds}{2\pi i} \; e^{st}\; \tilde{\cal A}_{Ti}(\vec k, s)\nonumber\\
 =&&\int_{c-i\infty}^{c+i\infty}\frac{ds}{2\pi i}\;e^{st}\;
\frac{s\, {\cal A}_{Ti}(\vec{k},0)+\dot{\cal A}_{Ti}(\vec{k},0)+\tilde{\cal
J}_{Ti}(\vec k,s)}  
{s^2+k^2+\Sigma^t(s)}\; .\label{Lapinverse} 
\end{eqnarray} 
Here we choose $ c \geq 0 $, such that the contour is  to the
right of all the singularities in the $s$-plane.

The time dependence of this integral crucially  depends on the
analytic properties of the  
propagator and hence a clear understanding of the poles and cuts of the
retarded propagator is essential.  At this stage we will set the
external current to zero, and analyze the contribution of the sources
at the end of this subsection.  
  
In all of the above and in what follows it is implicitly assumed that
the zero temperature   
divergences have been dealt with already by renormalizing the
amplitude of the field, i.e. wave-function renormalization and that we
are working with the  
subtracted spectral function. Recall that  the divergences are determined
solely by zero temperature fluctuations. 
%the contribution from Landau
%damping does not require any subtractions.  

We define the  
real and imaginary parts of the self-energy near the imaginary axis through 
\begin{equation} 
\Sigma^t(i\omega\pm0^+)=\Sigma^t_R(i\omega\pm0^+)+i\Sigma^t_I(i\omega\pm0^+). 
\end{equation} 

In fact it is easy to see from Eq.(\ref{selfenergy}) that 
\begin{eqnarray} 
\Sigma^t_I(i\omega\pm0^+)=&& 
\mp\text{sgn}(\omega)\frac{\pi}{2} 
(\rho_{ij}(|\omega|)-\rho_{ij}(-|\omega|))\label{specrepre}\; \\\nonumber 
=&&\pm \frac{e^2}{8\pi^2} \text{sgn}(\omega) 
\int\frac{d^3p\; p^2 \sin^2\theta}{\omega_p\omega_{k+p}} 
\left[ 
(1+n_p+n_{k+p})\delta(\omega_{k+p}+\omega_p-|\omega|)+\right.\\\nonumber 
&&\left.+(n_p-n_{k+p})\left\{\delta(\omega_{k+p}-\omega_p-|\omega|) 
-\delta(\omega_{k+p}-\omega_p+|\omega|)\right\}\right]. 
\end{eqnarray} 
 
This shows that $\Sigma^t_I(i\omega-0^+)=-\Sigma^t_I(i\omega+0^+)$ and
that the cuts will appear whenever the delta functions are
satisfied. Analysis of the arguments of the delta functions reveals
that the two-particle cuts arising from the first term in Eq.(\ref{specrepre})
stretch from $ s=\pm i(m+\omega_k) $ to $ s=
\pm i\infty $.  The second and third terms in Eq.(\ref{specrepre})
contain the hard thermal contributions. They   have
support when $\omega_{k+p}-\omega_p\simeq  
\vec{k}\cdot\hat{p}=\pm \omega$ which corresponds to the four-vector 
$(\omega,\vec{k})$ being spacelike. The resulting branch cut runs from $s=-ik$ 
to $s=+ik$. Hence these processes are induced only by  off-shell
space-like photons.  These HTL contributions lead to {\bf Landau
damping}. In summary, the 
physical region singularities and the Landau discontinuities show up as 
discontinuities in the imaginary part of the self-energy $\Sigma^t(s)$ when 
approaching the imaginary axis of the $s$-plane i.e. $s=i\omega$. 

In addition,  as argued in the previous section the contribution of the
physical region cuts from $ s=\pm i(m+\omega_k) $ to $ s=  \pm i\infty $
to the self-energy are $\propto \ln T$ only, and 
therefore subleading compared with the $O(T^2)$ terms.

Furthermore the long time dynamics due to the cuts will be dominated
by the thresholds (or the end-points).
Since the end-points of the Landau damping discontinuities are at $s\approx \pm
ik$, they will be the dominant contribution  
and the two particle cuts  from $s=\pm i(m+\omega_k)$ to 
$s= \pm\infty$ will be subleading at long times. Thus we can
simply focus on Landau damping both as the leading high temperature and long
time contributions. This argument is necessary because it is not {\em a-priori}
obvious that the high temperature and long time limits are described by the
same processes.

The leading term 
in the self-energy $\propto T^2$ is given by, 
\begin{eqnarray} 
&&\Sigma^t_I(i\omega\pm0^+)=\\\nonumber 
&&\simeq\pm\frac{e^2}{8\pi} \text{sgn}(\omega) 
\int dp\; p^2(-\frac{dn_p}{dp})\int_{-1}^1dx\;(1-x^2)kx \;
[\delta(kx-|\omega|)-\delta(kx+|\omega|)]\\\nonumber 
&&=\pm\frac{e^2T^2\pi}{12}\frac{\omega}{k} 
\left(1-\frac{\omega^2}{k^2}\right)\Theta(k^2-\omega^2). 
\end{eqnarray} 
The real part can be obtained either by using dispersion relations or by 
explicitly solving for the hard thermal self-energy  by  calculating the relevant integrals and we find   
\begin{equation} 
\Sigma^t(s)=-\frac{e^2T^2}{12}\left[ 
2\frac{s^2}{k^2}+i\frac{s}{k}\left(1+\frac{s^2}{k^2} \right) 
\ln\left(\frac{is-k}{is+k}\right)\right]+{\cal O}(\ln T). 
\label{htlsigmat} 
\end{equation} 
From this expression the transverse self-energy along the imaginary axis when
approaching from the right, can be 
read off easily: 
\begin{eqnarray} 
&&\Sigma^t(i\omega+0^+)=\nonumber\\ 
&&\frac{e^2T^2}{12} 
\left[2\frac{\omega^2}{k^2}+\frac{\omega}{k} 
\left(1-\frac{\omega^2}{k^2} \right) 
\ln\left|\frac{k+\omega}{k-\omega}\right|\right] 
+i\frac{e^2T^2\pi}{12}\frac{\omega}{k} 
\left(1-\frac{\omega^2}{k^2}\right)\Theta(k^2-\omega^2). 
\label{tbelowcut} 
\end{eqnarray} 
and agrees with known results\cite{htlnonabel,weldon1,lebellac,rebhan,inmedium}.

In addition to the branch cut singularities, the retarded propagator will 
also have isolated poles corresponding to the quasiparticle excitations which
can propagate in the plasma. The poles for the transverse excitations will be
given by the solutions to  
\begin{eqnarray} 
\omega_P^2=&&k^2+\frac{e^2T^2}{12} 
\left[2\frac{\omega_P^2}{k^2}+\frac{\omega_P}{k} 
\left(1-\frac{\omega_P^2}{k^2} \right) 
\ln\left|\frac{k+\omega_P}{k-\omega_P}\right|\right]+\label{poletrans}\\ 
&&+i\frac{e^2T^2\pi}{12}\frac{\omega_P}{k} 
\left(1-\frac{\omega_P^2}{k^2}\right)\Theta(k^2-\omega_P^2). 
\end{eqnarray} 
They will be in the physical sheet provided the
imaginary part vanishes at the pole.   
 
Though the equation cannot be solved analytically, 
in the case of interest when 
the external momenta are extremely soft (for e.g. $k\sim e^2T\ll eT$) 
representing a small amplitude long wavelength field configuration, the 
approximate location of the poles is found to be $s=\pm 
i\omega_P\simeq\pm i\frac{eT}{3}$\cite{htlnonabel,weldon1,lebellac,rebhan}.

The two-particle cuts were shown to run from $s=\pm 
i(m+\omega_k)$ to $\pm i\infty$ where $m$ is the mass of the scalar. A
consistent HTL resummation should also include the shift in the scalar masses,
thus ensuring that to this order the quasiparticle pole is in the 
physical sheet. Higher order contributions will provide a collisional
broadening to the pole.
It is a noteworthy point that the quasiparticle poles are located  beyond 
the Landau  discontinuities which stretch from $-ik$ to $+ik$ in the $s$-plane,
and below the two-particle threshold.

In summary, to this order in the HTL approximation, the analytic structure of
the retarded propagator in the high temperature limit features a Landau   
discontinuity running from $-ik$ to $+ik$ and quasiparticle poles at $s=\pm 
i\omega_p(k)$. The two-particle cut contributions have been shown to give a
subleading contribution both in temperature and in the long time dynamics. 
 
Using Eq.(\ref{Lapinverse}) we can now invert the transform by deforming the 
contour and wrapping it around the poles and the cuts to pick up the 
corresponding residues and discontinuities repectively so that, 
\begin{eqnarray} 
{\cal A}_{Ti}(\vec k,t)={\cal A}_{Ti}^{pole}(\vec k,t) 
+{\cal A}_{Ti}^{cut}(\vec k,t),\label{tpolecut} 
\end{eqnarray} 
The contributions from the  
quasiparticle poles add up to give a purely oscillatory behaviour in time. The
residues at the poles give rise to a wave function renormalization
with both $T$-dependent and $T$-independent contributions. The $T$-independent
contribution contains the typical logarithmic divergence and has been absorbed
in the usual zero temperature wave-function renormalization. We thus obtain,  
\begin{eqnarray} 
&&{\cal A}_{Ti}^{pole}(\vec k,t)=
Z^t[T]\;\left[ {\cal A}_{Ti}(\vec k,0)   \cos(\omega_Pt)+ \dot{\cal A}_{Ti}(\vec k,0) \frac{\sin(\omega_Pt)}{\omega_P} \right]
,\label{tpoleonly}\\\nonumber\\ 
&&Z^t[T]=\left[1-\frac{\partial\Sigma^t(i\omega)}{\partial\omega^2}\right] 
_{\omega=\omega_P\simeq eT/3}^{-1} \label{zetatrans} \; .
\end{eqnarray} 
Here $Z^t(T)$ is the temperature dependent wave function 
renormalization defined  
on-shell at the quasiparticle pole whose leading HTL contribution is obtained
from the self-energy (\ref{tbelowcut}). The continuum contribution is given by 
 
\begin{eqnarray}
{\cal A}_{Ti}^{cut}(\vec k,t)=
\frac{2}{\pi}\; 
\int_0^k d\omega\; \frac{\Sigma^t_I(i\omega+0^+)\left[{\cal A}_{Ti}(\vec
k,0)\; \omega \; \cos(\omega t)+ 
\dot{\cal A}_{Ti}(\vec k,0) \; \sin(\omega t)\right]}
{\left[ \omega^2-k^2-\Sigma^t_R(i\omega)\right]^2+\left[
\Sigma^t_I(i\omega+0^+)\right]^2}\; .\label{tcutonly}
\end{eqnarray} 
Evaluating Eqs.(\ref{tpolecut}),(\ref{tpoleonly}) and (\ref{tcutonly})
at $t=0$ we obtain an important sum rule, 
\begin{eqnarray} 
Z^t[T]+\frac{2}{\pi}\int_0^kd\omega\frac{\omega\Sigma^t_I(i\omega+0^+)} 
{\left[\omega^2-k^2-\Sigma^t_R(i\omega)\right]^2+
\left[\Sigma^t_I(i\omega+0^+)\right]^2}=1
\label{sumrule} \; .
\end{eqnarray}

This sum rule is a consequence of the canonical commutation relations but its
content in the HTL approximation is that in the high temperature
limit the wave function renormalisation which is evaluated {\em on-shell} is
completely determined by the Landau discontinuities which originate
from {\em strongly  
off-shell} processes. A similar sum rule was also obtained in
\cite{rob1} using different methods. 

The integral over the cut  (\ref{tcutonly})
cannot be evaluated in closed form but its long time asymptotics is dominated
by the end-point contributions as can be understood from the following argument.
The integral along the real $\omega$ axis from $\omega =0$ to $\omega=k$ can be obtained by deforming the 
integral into the  upper complex $\omega$ plane so that it runs along $\omega =
iz\; ; 0<z<\infty$, then around an arc 
at infinity and back to the real axis along the line $\omega=k+iz \; ; \;
0<z<\infty$ for the term $\propto e^{i\omega t}$ 
and similarly  into the lower complex plane for the term $\propto e^{-i\omega
t}$.  
 A detailed analysis of the asymptotic behavior of the integral reveals that only the $\omega = k$ 
end-point contributes because the contributions from $ \omega = 0 $ vanish
for large $ t $ faster than any negative power of $ t $. This is a consequence of the regular behavior of the
spectral density in the vicinity of $\omega =0$. 
 
We find the contribution from the $ \omega = k $ end-point in the long time
limit $ t>>1/k $ to be given by,
\begin{equation}\label{extremo}
{\cal A}_{Ti}^{cut}(\vec k,t)\buildrel{t \to \infty}\over=-
 \frac{12}{e^2T^2}\left\{{\cal A}_{Ti}(\vec
 k,0)\;\frac{\cos(kt)}{t^2}+ \dot{\cal A}_{Ti}(\vec
 k,0)\;\frac{\sin(kt)}{kt^2}\right\} 
\left[1+{\cal O}\left(\frac{1}{t}\right)\right]\quad .
\end{equation} 

>From the sum rule (\ref{tcutonly}) and (\ref{sumrule}) we find that the
{\em early} time behavior is approximately given by 
\begin{equation}
{\cal A}_{Ti}(\vec k,t) = {\cal A}_{Ti}(\vec k,0)\left[1-e^2T^2 t^2
\Delta_e+ {\cal O}(t^4)  \right] + \dot{\cal A}_{Ti}(\vec k,0)\Delta_o
\frac{e^2T^2}{k^2}\; t \; [1+ {\cal O}(t^2)]\; ,
\end{equation}
where $\Delta_{e,o}$ are constants depending on the wave-function
renormalization and  moments of the spectral density.

%*****
At long times, the dominant contributions are from the nearest
singularities in the complex $\omega$ plane.  This includes
the usual contributions from the quasiparticle modes at
$\omega = \pm \omega_P$, given by Eq. (\ref{tpoleonly}).
In addition, there is also the contributions from the branch
points at $\omega = \pm k$, Eq.(\ref{extremo}).  
There is no contribution from $\omega = 0$, because the
HTL self energy is regular about zero frequency.
%*****

\begin{figure}[t] 
\epsfig{file=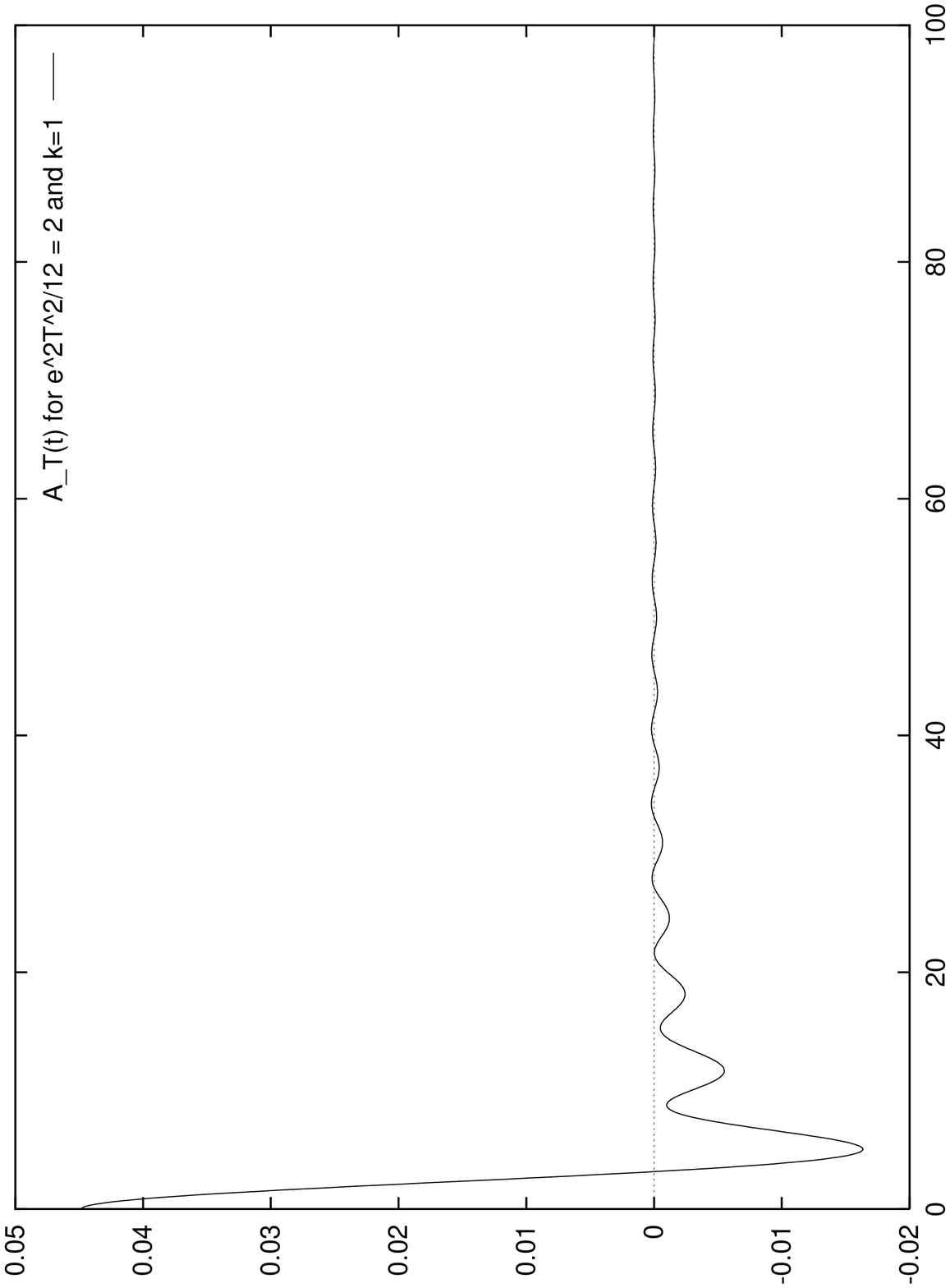,width=12cm,height=18cm} 
\caption{ Cut contribution  ${\cal A}^{cut}_T(k,t)/{\cal A}_T(k,0)$ for
$e^2T^2/12 = 2$ and $k=1$ vs. t. \label{fig1}} 
\end{figure}

\begin{figure}[t] 
\epsfig{file=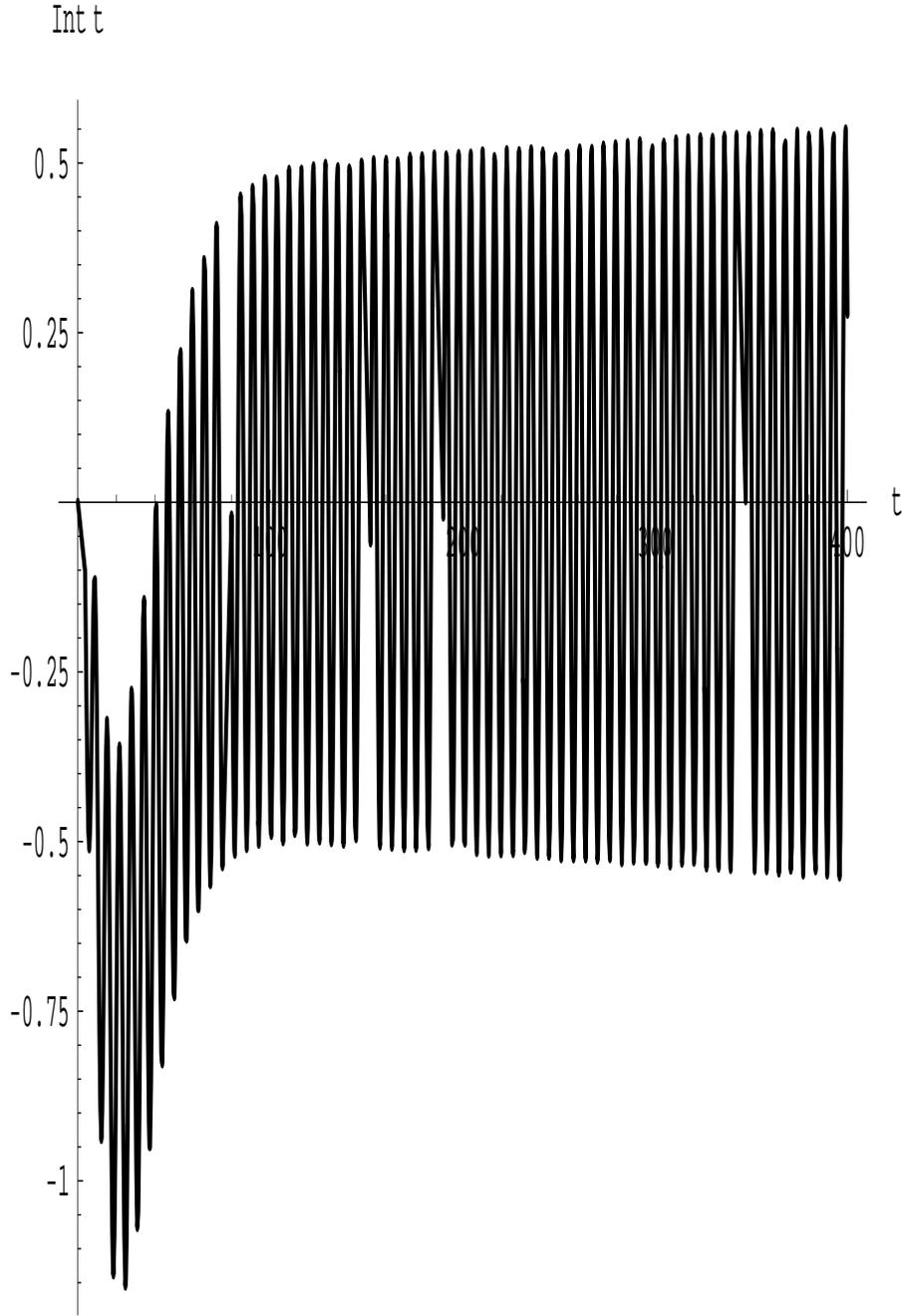,width=12cm,height=18cm} 
\caption{ $t^2 \times {\cal A}^{cut}_T(k,t)/{\cal A}_T(k,0)$ vs. t (in units of
1/k) for $m^2_D/k^2 = 
12; m^2_D=e^2T^2/3$\label{fig2}}  
\end{figure}

Figure (\ref{fig1}) shows the cut contribution ${\cal A}^{cut}_T(k,t)/{\cal
A}_T(k,0)$ vs. time for $e^2T^2/k^2=2$ and figure (\ref{fig2}) shows 
$t^2 \times {\cal A}^{cut}_T(k,t)/{\cal A}_T(k,0)$ vs. $ t $ (in units of
$ 1/k $) for $ m^2_D/k^2 =12 $ for the case $\dot{\cal A}_{Ti}(\vec k, 0)=0$.  

The point being made is that 
the  real time dynamics of the condensate is completely determined by the
analytic structure of the retarded propagator and the {\it global structure} of
the spectral density in the $s$-plane.    
 
The second important point to note is that the long time behaviour is
a {\bf power law} $\sim t^{-2}$ (times oscillations) and {\bf not an
exponential decay}. This means that Landau damping effects {\bf cannot} be
reproduced by phenomenological `viscous' terms  of the type
$\sim\Gamma \; \frac{d}{dt}$ neither at long nor at short times. The failure of
such a  phenomenologically motivated ansatz was already 
noticed at zero temperature in different contexts in ref.\cite{disip}. 
 We stress that such a
description not only fails to reproduce the power law behaviour but in fact
ignores {\it all the non-local} physics of Landau damping which is so clearly
encoded in the Hard Thermal Loop kernels. 

One might argue that higher order processes, both Landau damping and
collisional, could lead to an exponential relaxation. However, the point of the
above analysis is to argue that the full relaxational physics will be described
by a {\em competition} between the power laws from the lowest order Landau damping
contributions and the higher order exponential damping. The time scale of
interest will determine which process dominates.

For non-zero external sources we can obtain the general real time
evolution by inserting the Laplace transform of
the source in Eq.(\ref{Lapinverse}) 
$$
\tilde{\cal J}_{Ti}(\vec k,s)  =\int_0^\infty dt\; e^{-st} \;
{\cal J}_{Ti}(\vec{k},t) \; .
$$ 
This is an analytic function of $ s $ for $ Re(s) > 0 $ provided $
{\cal J}_{Ti}(\vec{k},t) $ is a non-singular function of
time. Let us assume that $ \tilde{\cal J}_{Ti}(\vec k,s) $ is analytic for
$ Re(s) > -\alpha $, where $ \alpha > 0 $ is a positive number. 
[Since $ \tilde{\cal J}_{Ti}(\vec k,s) $ vanishes for $  Re(s) \to
+\infty , \; \tilde{\cal J}_{Ti}(\vec k,s) $ must have singularities
somewhere in the left half-plane. Otherwise, it will be an 
entire function which is zero at infinity and therefore identically zero]. 
The singularities of  $ \tilde{\cal J}_{Ti}(\vec k,s) $ in the
left-half $s$ plane yield contributions to $ {\cal A}_{Ti}(\vec k, t)
$ through Eq.(\ref{Lapinverse}) which decrease exponentially in time
as $ e^{- \alpha t } $. They are therefore subdominant compared with
the Landau cut contributions (\ref{extremo}).

The source term contribution from the small $ s = i \omega$ region can be
understood simply as follows. Within the above hypothesis, the source term 
 $ \tilde{\cal J}_{Ti}(\vec k,i\omega ) $ can be expanded in a Taylor series
for small $\omega$  
$$
\tilde{\cal J}_{Ti}(\vec k,i\omega) = J_0(\vec k) + J_1(\vec k) \;
\omega + {\cal O}(\omega^2) \; .
$$

%*****
The even 
term, $ J_0(\vec k) $ can be absorbed into a shift of $ \dot{\cal A}_{Ti}(\vec
k,0) $, while the odd term $ J_1(\vec k) $ can 
be absorbed into a shift of $ {\cal A}_{Ti}(\vec k,0) $,
(\ref{Lapinverse}).  Thus as before the asymptotic large
time behavior is completely dominated
%*****
by the branch points at $\omega = \pm k$.
Under the above assumptions on the Laplace transform
 of the external current, we find the general time dependence to be given by 
\begin{eqnarray}
&&{\cal A}_{Ti}^{cut}(\vec k,t)\buildrel{t \to \infty}\over=\\\nonumber
&&-\frac{12}{e^2T^2\; kt^2 }\left\{ [ k {\cal A}_{Ti}(\vec k,0) - S_i(k) ]
\;\cos(kt) + [ \dot{\cal A}_{Ti}(\vec k,0) + C_i(k) ]\;
\sin(kt) \right\}\\\nonumber
&&\left[1+{\cal O}\left(\frac{1}{t}\right)\right]\; .
\end{eqnarray}
Here $ C_i(\omega) $ and  $ S_i(\omega) $ stand for the Fourier cosine
and sine transforms of the source ${\cal J}_{Ti}(\vec{k},t) $, respectively:
$$
 C_i(\omega) = \int_0^{\infty}  dt \; \cos(\omega t) \; {\cal
 J}_{Ti}(\vec{k},t) \; , \; S_i(\omega) = \int_0^{\infty} dt \;
 \sin(\omega t) \; {\cal  J}_{Ti}(\vec{k},t)\; .
$$

We emphasize that to this order in the HTL resummation, 
the region near $\omega \approx 0$ of 
the spectral density is regular (no branch singularities) 
and therefore {\em does not} contribute
 to the long time dynamics, which is completely determined 
by the end point (branch point) at $\omega =k$. 
%*********
%*********

\subsection{Longitudinal part} 

An analysis very similar to the one outlined above yields the following
expressions for the longitudinal component of the photon self-energy
in the case of an impulsive source ${\cal J}_L(\vec x,t) =
\delta^3(\vec x) \delta(t)$ with spatial Fourier and Laplace transform
given by $\tilde{\cal J}_L(\vec k,s) =1$. The case of 
a more complicated source can be obtained by convolution. In this case
we obtain 
\begin{equation}
{\cal A}_0(\vec k,s)=\frac{1}{k^2+\Sigma^l(s)}
\end{equation}
where  
\begin{equation} 
\Sigma^l(s)=\frac{e^2T^2}{3}-\frac{e^2T^2}{6} 
\frac{s}{ik}\;\ln\left(\frac{is-k}{is+k}
\right)\; .\label{htlsigmal} 
\end{equation} 
Thus the longitudinal self-energy along the imaginary axis in the $s$-plane, 
when approaching from the right is obtained as before to be: 
\begin{equation} 
\Sigma^l(i\omega+0^+)=\frac{e^2T^2}{3}\left[ 
1-\frac{\omega}{2k}\ln\left|\frac{k+\omega}{k-\omega}\right|
\right]-i\frac{e^2T^2\pi}{6}\frac{\omega}{k}\;
\Theta(k^2-\omega^2)\; . 
\label{lbelowcut} 
\end{equation} 
The location of the longitudinal quasiparticle or plasmon poles is
given by the following dispersion relation,
\begin{equation} 
k^2+\frac{e^2T^2}{3} 
\left[1-\frac{\omega}{2k} \ln\left|\frac{k+\omega}{k-\omega}\right|
\right]-i\frac{e^2T^2\pi}{6}\frac{\omega}{k} 
\Theta(k^2-\omega^2)=0\;. \label{polelongi}
\end{equation} 
For soft external momenta $(k<<eT)$ the plasmon poles can be seen to be at 
$s=\pm i\omega_0\simeq \pm ieT/3$. The real time dependence of the longitudinal 
condensate is then found by inverting the transform using 
\begin{eqnarray}
{\cal A}_0(\vec k,t)=\int_{c-i\infty}^{c+i\infty}\frac{ds}{2\pi i}\,
e^{st}\, \tilde {\cal A}_0(\vec k,s) 
\end{eqnarray}
where the contour is to the right of all the singularities as in 
Eq.(\ref{Lapinverse}). As in the 
transverse case, at high temperatures the singularity structure is
dominated by the discontinuity across the cut that runs from $-ik$ to $+ik$
corresponding to Landau damping,  and the two plasmon poles at  
$\pm i\omega_0$. The Bromwich contour is then deformed to pick up the cut
and pole contributions so that
\begin{eqnarray} 
{\cal A}_{0}(\vec k,t)={\cal A}_{0}^{pole}(\vec k,t)
+{\cal A}_{0}^{cut}(\vec k,t),\label{lpolecut} 
\end{eqnarray} 
where 
\begin{eqnarray} 
&&{\cal A}_{0}^{pole}(\vec k,t)=
-Z^l[T]\;\frac{\sin(\omega_0t)}{\omega_0},\label{lpoleonly}\\\nonumber\\ 
&&Z^l[T]=\left[\frac{\partial\Sigma^l(i\omega)}{\partial\omega^2}\right] 
_{\omega=\omega_o\simeq eT/3}^{-1} \label{zetalongi} 
\end{eqnarray} 
and 
\begin{eqnarray} 
{\cal A}_{0}^{cut}(\vec k,t)=
-\frac{2}{\pi}\;\int_0^kd\omega\frac{\Sigma^l_I(i\omega+0^+)\sin(\omega t)} 
{\left[k^2+\Sigma^l_R(i\omega)\right]^2+\left[\Sigma^l_I(i\omega+0^+)
\right]^2}
\;. \label{lcutonly}
\end{eqnarray}
Unlike the transverse components for which the sum rule is a
consequence of the
canonical commutation relations, for the longitudinal component there
is no equivalent sum rule because the field $A_0(x)$ is a non-propagating
Lagrange multiplier.

The long time, asymptotic behaviour of the longitudinal condensate is
dominated by the end-points of the integral (\ref{lcutonly}). The
end-point $ \omega = 0 $ yields contributions that vanish for long $ t
$ faster than any negative power of $ t $, as it was the case for the
transverse part (\ref{tcutonly}). We find that the  long time asymptotics
 for $t>>1/k $ is dominated by  the  end-point contribution at 
$ \omega = k $,
\begin{eqnarray}\label{asiP}
{\cal A}_0^{cut}(\vec k,t)&=&
a_{asymp}^{cut}(\vec k,t)\left[1+{\cal
O}\left(\frac{1}{t}\right)\right] \;\;, \cr \cr
a_{asymp}^{cut}(\vec k,t)&\equiv &-{12 \over {e^2T^2}}\;
\int_0^{\infty} dx \, e^{-x} \; {{\cos[ k t + \alpha(x,t) ]}
\over {\sqrt{\log^4{{cx}\over {kt}} + {{5\pi^2}\over 2} \;
\log^2{{cx}\over {kt}} + {{9 \pi^4}\over {16}} }}} \label{asiL}
 \\ 
\alpha(x,t)& \equiv & \arctan{{\pi \log{{cx}\over {kt}}}\over 
{ {{3\pi^2}\over 4} + \log^2{{cx}\over {kt}}}}  \label{alfalongi}\\
c \; & \equiv & \frac12 \; \exp[ 2 + {{6k^2}\over {e^2T^2}}]\; , \label{clongi}
\end{eqnarray}

The integral in Eq.(\ref{asiL}) cannot be expressed in terms of
elementary functions and it is related to the $ \nu(x) $ function
\cite{erde}. For large $ t $, one can derive an asymptotic expansion
in inverse powers of $ \log [kt/c] $ by integrating Eq.(\ref{asiL}) by parts:
\begin{eqnarray}
a_{asymp}^{cut}(\vec k,t) & \buildrel{t \to \infty}\over= & 
-{12 \over {e^2T^2\; t}}\; {{\cos(kt + \beta(kt))}\over{{\log^2{{kt}\over
c} +  {{\pi^2}\over 4} }}} \left[1+{\cal
O}\left(\frac{1}{\log t}\right)\right] \label{asiLD}\\
\beta(kt) & \equiv &  -\arctan{{\pi \log{kt}}\over 
{\log^2{kt} -{{\pi^2}\over 4}}} \; ,\label{betalongi}
\end{eqnarray}

\begin{figure}[t] 
\epsfig{file=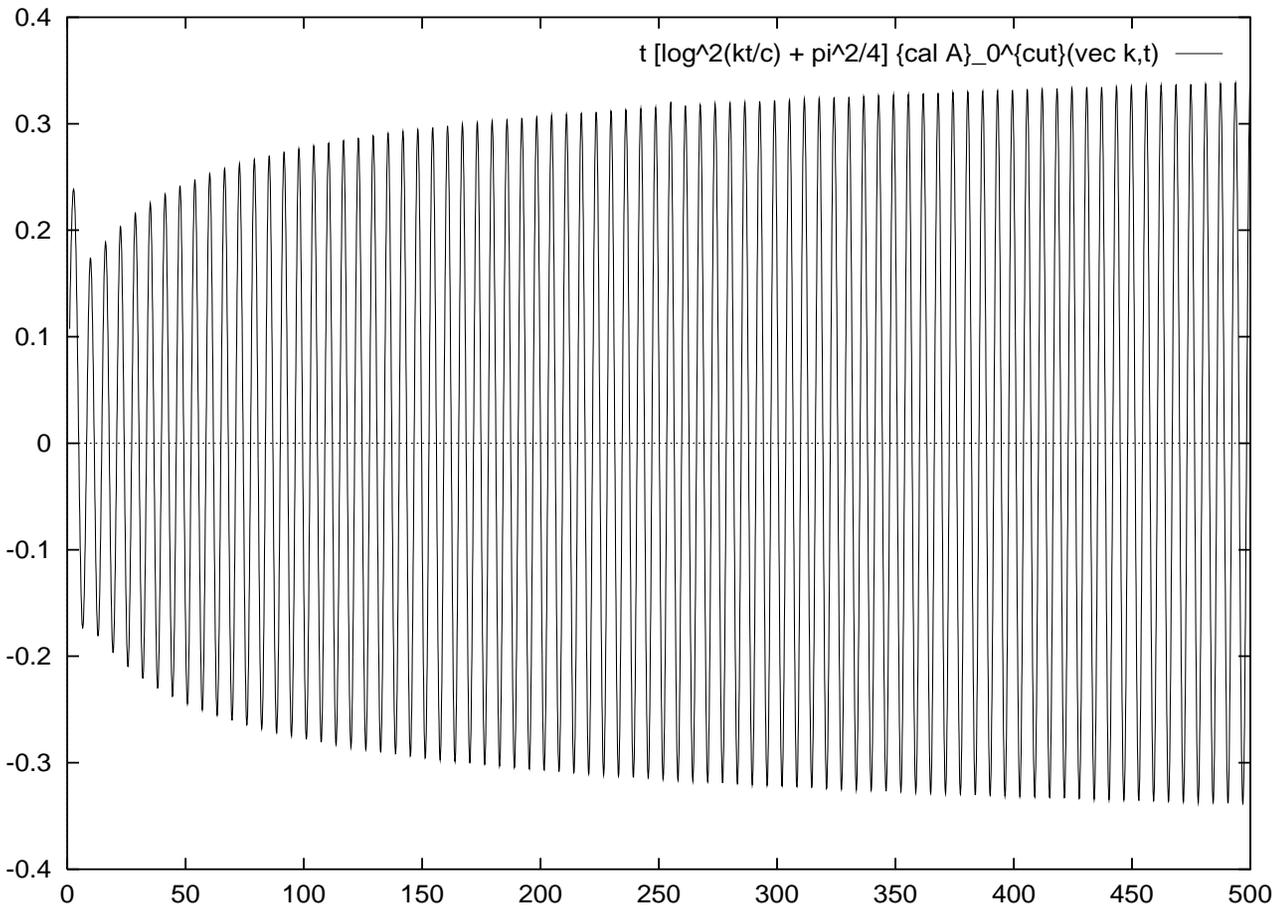,width=12cm,height=18cm} 
\caption{ $ t [\log^2(kt/c) + \pi^2/4] {\cal A}_0^{cut}({\vec k},t) $
as a function of $ t $ for $ e^2 T^2 = 6 $ and $ k = 1 $
[see Eqs.(\ref{asiL})-(\ref{asiLD})]. \label{fig3a} }
\end{figure}

\begin{figure}[t] 
\epsfig{file=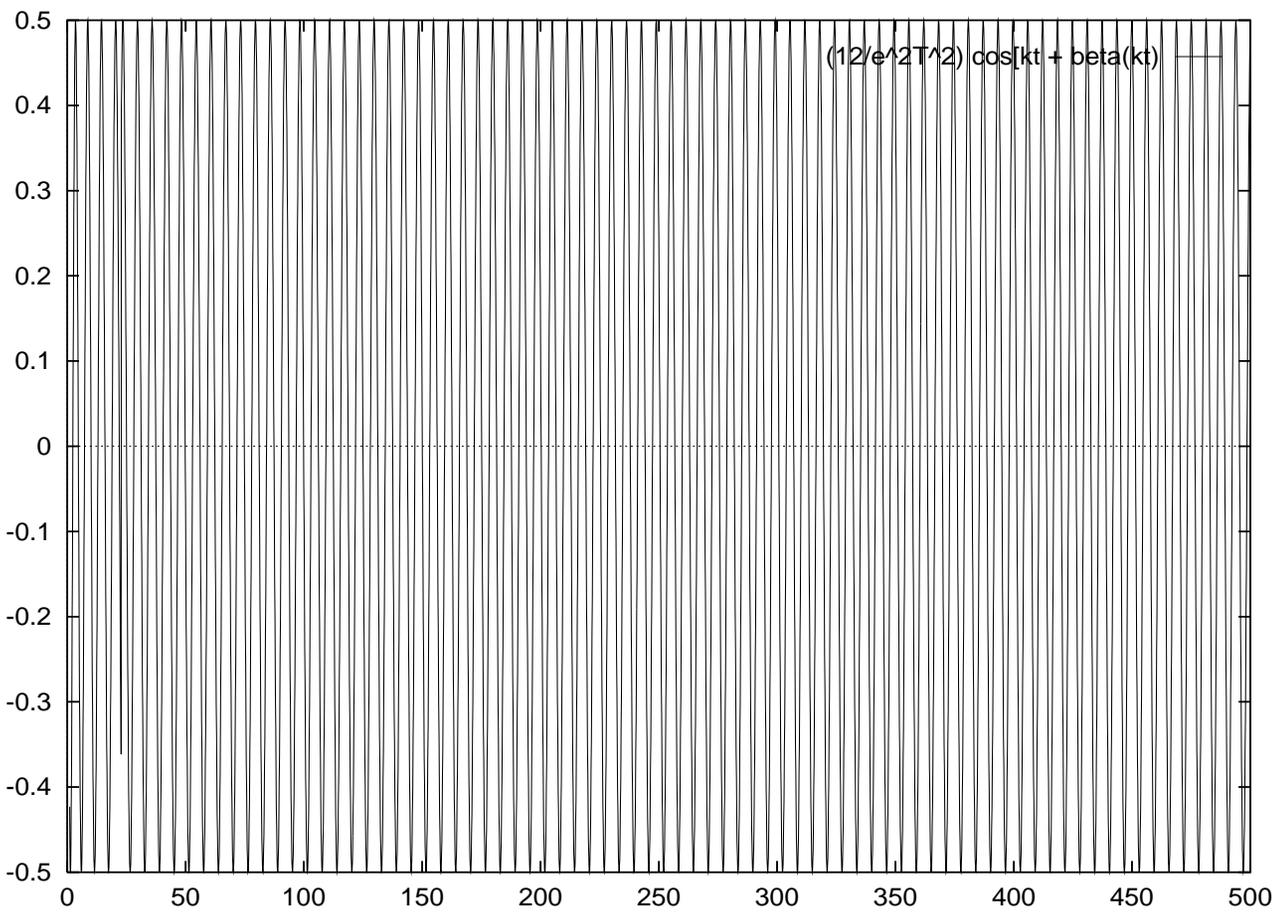,width=12cm,height=18cm} 
\caption{$(12/e^2 T^2 ) \cos[ kt + \beta(kt) ] $  a function of $ t $ for 
$ e^2 T^2 = 6 $ and $ k = 1 $ [see Eqs.(\ref{asiL})-(\ref{asiLD})]. 
\label{fig3b}}
\end{figure}

This expansion is not very good quantitatively unless $ t $ is very
large. For example, for $ kt = 500 ,\;  a_{asymp}^{cut}(\vec
k,t) $ given by Eq.(\ref{asiP})
approximates $ {\cal A}_0^{cut}(\vec k,t) $ up to $ 0.1 \% $,
whereas, the dominant term in Eq.(\ref{asiLD}) is about $ 30\% $
smaller than $ a_{asymp}^{cut}(\vec k,t) $. Figure (\ref{fig3a}) shows
$ t [\log^2(kt/c) + \pi^2/4] {\cal A}_0^{cut}({\vec k},t) $ and
(\ref{fig3b}) shows 
$(12/e^2 T^2 ) \cos[ kt + \beta(kt) ] $, (see
Eqs.(\ref{asiL})-(\ref{asiLD})), both figures should coincide 
for large enough $ t $ but we see
numerically a sizable discrepancy even for $ t = 500 $ due to the slow
convergence of the  asymptotic expansion (\ref{asiLD}).   

\begin{figure}[t] 
\epsfig{file=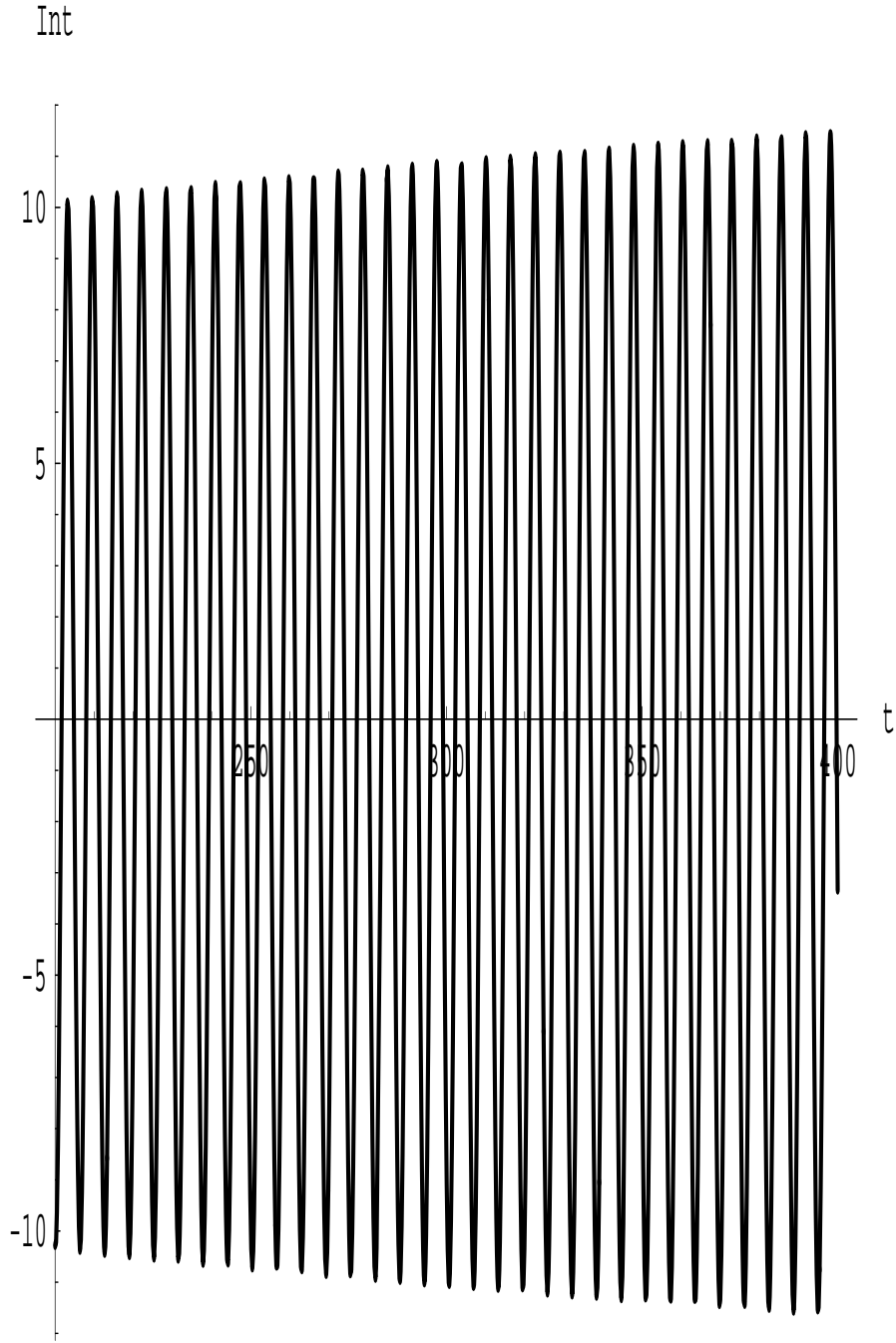,width=12cm,height=18cm} 
\caption{ $t \times (\ln[t])^{2.5}\times {\cal A}^{cut}_0(k,t)$ vs. t (in units
of 1/k) for $m^2_D/k^2 = 2$. \label{fig4}}  
\end{figure}

A saddle point analysis yields an intermediate asymptotics of the form
\begin{equation}
a_{asymp}^{cut}(\vec k,t) \approx
\frac{3\sqrt\pi\cos(kt)}{e^2T^2t\ln^{2.5}[kt]}
\label{saddleptlongi}
\end{equation}
which although is a relatively good estimate within a time window,
(see fig. (\ref{fig4})) is seen numerically to have a discrepancy of the same
order as the dominant term above in a wider range of time.  
Nevertheless, the exact expression (\ref{asiL}) is easily obtained numerically
for arbitrary range of parameters. 

The case of a general source requires a convolution. The cut
contribution takes now the form,

\begin{eqnarray} \label{ALfuente}
{\cal A}_{0}^{cut}(\vec k,t)=
-\frac{2}{\pi}\;\int_0^kd\omega\frac{\Sigma^l_I(i\omega+0^+)\left[ 
C_L(\omega) \;\sin(\omega t) - S_L(\omega) \;\cos(\omega t) \right]}
{\left[k^2+\Sigma^l_R(i\omega)\right]^2+\left[\Sigma^l_I(i\omega+0^+)
\right]^2} \;. \nonumber
\end{eqnarray}

where $ C_L(\omega) $ and $ S_L(\omega) $ stand for the cosine
and sine Fourier coefficients of the source ${\cal J}_{L}(\vec{k},t) $,
respectively. 

For a longitudinal source fullfiling the same general analyticity assumptions
as 
the transverse source we can  expand  in series for small $\omega$ as 
$$
\tilde{\cal J}_L(\vec k, i\omega) = J_{0L}+ J_{1L}\; \omega + {\cal
O}(\omega^2) \; . 
$$ 

As for Eq.(\ref{lcutonly}) the end-point $ \omega = 0 $ of the
integral (\ref{ALfuente}) yields  contributions that vanish for long $ t$ 
faster than any negative power of $ t $. Just as in the transverse case, this result is a consequence of the
fact that to this order in the HTL resummation, the spectral density is 
regular (no branch points) near
$\omega =0$. 
%*****
%*****

To summarize, we gather
the final results for the asymptotic real-time evolution of the transverse
and longitudinal non-equilibrium expectation value in the linear approximation
\begin{itemize}
\item{{\bf Transverse (no external source):} 
\begin{eqnarray}
{\cal A}_{Ti}(\vec k,t) & = &  {\cal A}_{Ti}(\vec k,0) \left[
Z^t[T] \cos[\omega_pt] - \frac{12}{e^2T^2}\frac{\cos(kt)}{t^2}
+{\cal O}\left(\frac{1}{t^3}\right)\right] \nonumber \\
 & + & \dot{\cal A}_{Ti}(\vec k,0) \left[
Z^t[T] \frac{\sin[\omega_pt]}{\omega_p} - \frac{12}{e^2T^2}\frac{\sin(kt)}{kt^2}
+{\cal O}\left(\frac{1}{t^3}\right)\right]
\label{transversefinal} 
\end{eqnarray}
The sum rule (\ref{sumrule}) implies that the coherent field configuration
relaxes to an  asymptotic amplitude which is {\em smaller} than the initial, 
and the ratio of the final to the initial amplitude is completely determined by
the thermal wave function renormalization.     
}

\item{{\bf Longitudinal (impulsive external source): }
\begin{equation}
{\cal A}_{0}(\vec k,t) =   
-Z^l[T]\;\frac{\sin(\omega_0t)}{\omega_0}+
 a_{asymp}^{cut}(\vec k,t)\left[ 1 + {\cal O}\left(\frac{1}{t}\right)\right].
\end{equation}
}

The transverse and longitudinal wave function renormalizations
$Z^t[T]\;, Z^l[T]$ are given by Eqs. (\ref{zetatrans}) and (\ref{zetalongi})
respectively 
and $a_{asymp}^{cut}(\vec k,t)$ is given by Eqs. (\ref{asiL},\ref{asiLD}) while
the 
positions of the poles $\omega_p \; ,\; \omega_0$ can be obtained by the
solving  
Eqs. (\ref{poletrans}) and (\ref{polelongi}) respectively. 

%%%ADDITION 7 HERE

\item{{\bf In Summary:} we find that the long time dynamics is dominated by the Landau damping thresholds at $\omega = \pm k$, not by the $\omega \approx 0$ region of the spectral density. The early time dynamics is
determined by  moments of the total spectral density. 
%new paragraph%%%%%
Long time power law tails had recently found in the current correlators in reference\cite{yaff}}.
\end{itemize}

\section{Langevin description for the soft modes and Fluctuation-Dissipation
relation}  

\subsection{Langevin Description}

So far we have studied the damping of soft modes in the plasma by the hard 
particles from the microscopic point of view. 
%!!! A CHANGE HERE
In this section we provide a  stochastic description of the
relaxation of gauge fields via a semiclassical Langevin equation with
a Markovian damping kernel and a gaussian white noise.

A semiclassical description
treats the hard modes as a `bath' and the soft modes as the 
`system'. The bath degrees of freedom are integrated out, their main
effect being encoded in a dissipative kernel and a stochastic noise
inhomogeneity in the 
resulting Langevin equation. The dissipative kernel is related to the
stochastic correlation function of the noise via a generalized
fluctuation-dissipation relation.    
Physically the stochasticity arises because the 
hard scales which are integrated out in the HTL scheme and are 
responsible for Landau damping  will also provide random kicks to the soft
degrees of freedom.  
 
 This section is devoted to a {\it microscopic 
derivation} of the Langevin equation for the inhomogeneous gauge field 
configuration, to leading order in the hard thermal loop program.
This is achieved explicitly by integrating out the 
hard modes which provide a natural realization of the `bath' variables while 
the soft modes are to be treated as the `system' variables.  The methodology of
this approach is based on the Feynman-Vernon influence functional\cite{feynman}
which has already been used to describe dissipation and decoherence in quantum
systems from a microscopic theory\cite{disip,leggett,langevin}. 
 
The non-equilibrium partition function for the full field theory is  
\begin{eqnarray} 
{\cal Z}&=&\int{\cal D}A^{(\pm)}_{Ti}\; {\cal D}\Phi^{(\pm)}\;{\cal 
D}\Phi^{(\pm)\dagger}\;\exp\left[i\int
d^4x\left\{\partial_\mu\Phi^{(+)\dagger}\;
\partial^\mu\Phi^{(+)}\right.\right.\\\nonumber 
&+& \left.\left. 
\frac{1}{2}\partial_\mu\vec{A}_T^{(+)}\cdot\partial^\mu\vec{A}_T^{(+)}
-e\vec{A}_T^{(+)}\cdot\vec{j}_T^{(+)}-e^2\vec{A}_T^{(+)}\cdot\vec{A}_T^{(+)} 
\; \Phi^{(+)\dagger}\;\Phi^{(+)})-
\left[+\rightarrow-\right]\right\}\right]\\\nonumber
\end{eqnarray} 
and the effective action for the `system' (the soft photons) follows 
by performing  
the path integral over the `bath'  (the hard scalars) treating the `system' 
degrees of freedom  
as background fields. This means that all subsequent expectation
values will be  
evaluated in the reduced density matrix which defines the effective field 
theory for the system degrees of freedom.  
It is then convenient to introduce the `center of mass' and `relative' 
coordinates

\begin{equation} 
A_{Ti}^{(\pm)}={\cal A}_{Ti}\pm\frac{R_{Ti}}{2}. 
\end{equation} 
In terms of these redefined fields the effective action can be obtained to 
one-loop order via a systematic loop-expansion of the reduced partition 
function:   
\begin{eqnarray} 
S_{eff}[{\cal A}_{Ti},R_{Ti}]&&=\int d^4x \left\{ 
-R_{Ti}\;\partial^2{\cal A}_{Ti}-2e^2\;R_{Ti}\;{\cal A}_{Ti} \;
\langle\Phi^{\dagger}\Phi\rangle+\right.\\\nonumber 
&&\left.+\frac{ie^2}{4}\int d^4x^{\prime}R_{Ti}(x)\;R_{Tj}\;
\left[\langle j_{Ti}^{(+)}(x)j_{Tj}^{(-)}(x^\prime)\rangle+ 
\langle j_{Ti}^{(-)}(x)j_{Tj}^{(-)}(x^\prime)\rangle\right]\;
R_{Tj}(x^\prime)
\right. 
\\\nonumber 
&&\left.+ie^2\int  d^4x^{\prime}R_{Ti}(x)\left[ 
\langle j_{Ti}^{(-)}(x)j_{Tj}^{(+)}(x^\prime)\rangle-\langle 
j_{Ti}^{(+)}(x)j_{Tj}^{(-)}(x^\prime)\rangle\right]\Theta(x_0-x_0^\prime)\; 
{\cal A}_{Tj}(x^\prime)\right\} 
\end{eqnarray} 
The transverse current $j_{Ti}(x)$ was introduced in
Eq.(\ref{current}) and the  
curent-current correlators can be calculated easily using the defining 
formulas for the free scalar propagators in Eq.(\ref{gplpl}-\ref{lesser}).  
In order to make explicit the soft momentum scales of interest we perform
a spatial  Fourier transform, in terms of which the reduced effective
action, including the influence functional of the hard modes is given by 

\begin{eqnarray} 
S_{eff}[{\cal A}_{Ti},R_{Ti}]=&&\int \frac{d^3k}{(2\pi)^3}\int dt\left\{ 
-R_{Ti}(\vec k,t)\left(\frac{d^2}{dt^2}+k^2+2e^2\langle\Phi^{\dagger} 
\Phi\rangle\right){\cal A}_{Ti}(-\vec k,t)+\right. 
\label{effaction}\\\nonumber 
&&\left.-\int^tdt^\prime R_{Ti}(\vec k,t){\cal D}_{ij}(k;t,t^\prime) 
{\cal A}_{Tj}(-\vec k,t^\prime)+\right.\\\nonumber 
&&\left.+i\int dt^\prime R_{Ti}(\vec k,t){\cal N}_{ij}(k;t,t^\prime) 
R_{Tj}(-\vec k,t^\prime)\right\}  
\end{eqnarray} 
where ${\cal D}_{ij}$ and ${\cal N}_{ij}$ will be shown to be the dissipation 
and noise kernels respectively. 
The dissipation kernel which is given by  
\begin{eqnarray} 
&&{\cal D}_{ij}(\vec k;t,t^\prime)=\\\nonumber 
&&4ie^2\int\frac{d^3p}{(2\pi)^3} \;
p_{Ti}\;p_{Tj}\;[G_p^<(t,t^\prime)G_{k+p}^<(t,t^\prime)- 
G_p^>(t,t^\prime)G_{k+p}^>(t,t^\prime)]\Theta(t-t^\prime)\\\nonumber 
=&&2e^2\int \frac{d^3p}{(2\pi)^3\omega_p\omega_{k+p}}p_{Ti}p_{Tj} 
\left\{(1+n_p+n_{p+k})\sin[(\omega_p+\omega_{k+p})(t-t^\prime)]+ 
\right.\\\nonumber 
&&\left.+(n_p-n_{p+k})\sin[(\omega_{k+p}-\omega_p)(t-t^\prime)]\right\} 
\Theta(t-t^\prime) 
\end{eqnarray} 
clearly gives a causal contribution to the effective action as seen in 
Eq.(\ref{effaction}). The fact that it is real follows from the properties of 
the non-equilibrium Green's functions  
[Eqs.(\ref{greater}) and (\ref{lesser})] which imply that  
$$
G_p^<(t,t^\prime)G_{k+p}^<(t,t^\prime) 
-G_p^>(t,t^\prime)G_{k+p}^>(t,t^\prime)\sim 
2iIm\left[G_p^<(t,t^\prime)G_{k+p}^<(t,t^\prime)\right]\;.
$$
Furthermore the dissipation kernel is in fact precisely the one-loop
self-energy which appears
in the effective action equation of motion Eq.(\ref{transeqmotion1}). 
The noise kernel on the other hand is {\it acausal} and gives an {\it
imaginary}  contribution  to the effective action 
 
\begin{eqnarray} 
&&{\cal N}_{ij}(\vec k;t,t^\prime)=\\\nonumber 
&&-e^2\int\frac{d^3p}{(2\pi)^3} 
\; p_{Ti}\; p_{Tj}\; [G_p^<(t,t^\prime)G_{k+p}^<(t,t^\prime)+ 
G_p^>(t,t^\prime)G_{k+p}^>(t,t^\prime)]\\\nonumber 
=&&\frac{e^2}{2}\int \frac{d^3p}{(2\pi)^3\omega_p\omega_{k+p}}\;
p_{Ti}\; p_{Tj}\; 
\left\{(1+n_p+n_{p+k}+2n_pn_{p+k})\cos[(\omega_p+\omega_{k+p})(t-t^\prime)]+ 
\right.\\\nonumber 
&&\left.+(n_p+n_{p+k}+2n_pn_{p+k})\cos[(\omega_{k+p}-\omega_p)(t-t^\prime)] 
\right\}\; .
\end{eqnarray} 
Therefore, 
\begin{equation}
{\cal N}_{ij}\propto Re\left[\int d^3p\;p_{Ti}\; p_{Tj}\; \; G_p^<(t,t^\prime)
G_{k+p}^<(t,t^\prime)\right] 
\end{equation} 
which means that the noise-noise correlator and 
the dissipation kernel are the real and imaginary parts respectively
of the same  
analytic function of $(t-t^\prime)$. Thus they are automatically related by 
dispersion relations which reveal the fluctuation-dissipation theorem within 
this context. 
 
The imaginary, non-local, acausal part of the effective action gives a 
contribution to the path-integral that may be rewritten in terms of a
stochastic  field as\cite{langevin} 
 
\begin{eqnarray} 
{\cal Z}&=&\int {\cal D} R_{Ti}(\vec k)\; {\cal D} R_{Ti}^*(\vec k) 
\exp\left[-\int dt\; dt^\prime \frac{d^3k}{(2\pi)^3}\; R_{Ti}^*(\vec k,t) \;
{\cal N}_{ij}(k;t,t^\prime)\;R_{Tj}(\vec k,t^\prime)\right]\\\nonumber 
&\propto&\int{\cal D} R_{Ti}(\vec k)\;{\cal D} R_{Ti}^*(\vec k) \;
{\cal D} \xi_{i}(\vec k)\;{\cal D} \xi_{i}^*(\vec k)\;P[\xi] \;
\exp\left[i\int dt\frac{d^3k}{(2\pi)^3}\;\xi_i(\vec k,t)\;R_{Ti}(-\vec
k,t)+c.c\right]  
\end{eqnarray} 
where the noise has a Gaussian probability distribution  
\begin{equation} 
P[\xi]=\exp[-\int dt\;  dt^\prime\; \xi_i(\vec k,t) \;
{\cal N}^{-1}_{ij}(k;t,t^\prime)\;\xi_j(-\vec k,t^\prime)] \label{proba}
\end{equation} 
with zero mean and non-Markovian correlations: 
\begin{equation} 
<<\xi_i(\vec k, t)>>=0\quad ;\;\;\;\;<<\xi_i(\vec k,t) \;
\xi_j(\vec{k}^\prime,t^\prime)>>= 
(2\pi)^3\delta^{(3)}(\vec k +\vec{k}^\prime) \;
{\cal N}_{ij}(\vec k;t,t^\prime). \label{corre}
\end{equation}
where $<<\cdots>>$ is the {\it stochastic} average with the probability
distribution $P[\xi]$. 

In the Hard Thermal Loop approximation the leading terms in the noise kernel 
can  be calculated explicitly, giving 
 
\begin{eqnarray} 
{\cal N}_{ij}(\vec k;t,t^\prime)\simeq 
\frac{e^2T^3}{6}{\cal P}_{ij}(\vec k) 
\left[\frac{\sin[k(t-t^\prime)]}{k^3(t-t^\prime)^3} 
-\frac{\cos[k(t-t^\prime)]}{k^2(t-t^\prime)^2} 
\right] 
\end{eqnarray} 
 
Clearly these are {\it long range} correlations which cannot be replaced by  
local delta function in any regime of approximation. In fact the Fourier 
transform of the kernel yields: 
 
\begin{eqnarray} 
\tilde{\cal N}_{ij}(\vec k,w)&=&\int_{-\infty}^{+\infty}d(t-t')\; 
{\cal N}_{ij}(\vec k; 
t-t^\prime)\; e^{-i\omega(t-t^\prime)}\\\nonumber\\ 
=&&{\cal P}_{ij}(\vec k)\; \frac{e^2T^3}{24k}\left(1-\frac{\omega^2}{k^2} 
\right)\Theta(k^2-\omega^2)={\cal P}_{ij}(\vec k) \;
\frac{T}{2\omega}\;\Sigma_I(i\omega+0^+). 
\end{eqnarray} 
where $\Sigma_I(i\omega+0^+)$ is the imaginary part of the self-energy
given by  Eq.(\ref{tbelowcut}) resposible of the  Landau damping processes.

 This clearly shows Landau damping to be the 
origin of the noise correlation and also provides an explicit realization of 
the fluctuation-dissipation theorem. We draw attention to the factor
of $T/\omega$  which arises from the  high $T$ limit of the Bose
factor $(1+2n_\omega)$ (see below) and leads to a noise-noise correlation
$\propto e^2T^3$. This result is in accord with the usual
fluctuation-dissipation relation, in which the noise-noise correlation 
function has one more power of $T$ than the dissipative contribution to the
Langevin equation.  
 
The Langevin equation itself is obtained via the saddle point condition\cite{langevin} 
 
\begin{equation} 
\frac{\delta S_{eff}}{\delta R_i(\vec k,t)}|_{R=0}=\xi_i(\vec k,t) 
\end{equation} 
leading to 
\begin{equation} 
\frac{d^2}{dt^2}{\cal A}_{Ti}(\vec{k},t)+k^2{\cal 
A}_{Ti}(\vec{k},t)+2e^2\langle\Phi^\dagger\Phi\rangle{\cal A}_{Ti}(\vec{k},t) 
-\int_{-\infty}^tdt^\prime{\cal D}_{ij}(k;t,t^\prime) 
{\cal A}_{Tj}(\vec{k},t^\prime) = \xi_i(\vec k, t^\prime) \label{langevin}
\end{equation}

\subsection{Fluctuation-Dissipation relation:} 
The general form of the fluctuation-dissipation relation can be established at
this level by retaining the complete expressions for the dissipative and noise
kernels without performing the HTL approximation. A standard analysis of the spectral representation\cite{baymkad,fetter}
of the equilibrium correlators leads to the following result for 
the Fourier transform of the noise-noise correlation function 
\begin{equation}
\tilde{\cal N}_{ij}(\vec k,\omega) = \frac{1}{4} \;
\mbox{Im}\left[\Sigma^{ret}_{ij}(\vec k;\omega) \right]\coth\left[\frac{\beta
\omega}{2}\right]\label{noiseFT}  
\end{equation}
Finally taking temporal Fourier transforms of both sides of the Langevin
equation (\ref{langevin}) and averaging over the noise with the Gaussian distribution function  we find 
\begin{eqnarray}
&& <<\tilde{{\cal A}}_{Ti}(\vec k, \omega)\tilde{{\cal A}}_{Ti}(-\vec k,
-\omega)>> = |\tilde{{\cal A}}_{Ti}^H(\vec k, \omega)|^2+\rho(\vec
k,\omega)\coth\left[\frac{\beta \omega}{2}\right] \nonumber \\ 
&& \rho(\vec k,\omega) = \frac{1}{4}\frac{\mbox{Im}\left[\Sigma^{ret}_{ij}(\vec k,
\omega)\right] }{\left( \omega^2-\omega^2_k-\mbox{Re}\left[
\Sigma^{ret}_{ij}(\vec k,\omega)\right]\right)^2+\left(\mbox{Im}
\left[\Sigma^{ret}_{ij}( \vec k,\omega )\right]\right)^2} \label{flucdissi}  
\end{eqnarray} 
\noindent where the tildes stand for the Fourier transform and  $\tilde{{\cal A}}^H_{Ti}(\vec k, t)$ is a solution of the homogeneous
equation, which is precisely given by Eq. (\ref{transversefinal}) (assuming 
$\dot{{\cal A}}^H_{Ti}(\vec k, t=0)=0$ which can be relaxed with the proper
generalization). 
The double brackets stand for averages over the noise with the
probability distribution (\ref{proba})-(\ref{corre}). This is the form of the 
usual fluctuation-dissipation relation, which we obtained consistently by
integrating out 
the hard modes and deriving the influence functional\cite{feynman,leggett} for
the transverse components of the gauge invariant fields. The semiclassical
Langevin equation is useful in order to obtain semiclassical correlation
functions by averaging the solution of a partial differential equation over a
stochastic gaussian noise. The question arises: to which correlation function
of the microscopic theory are these stochastic averages related? The answer
to this question is found by writing a spectral representation of the
equilibrium correlator $\langle {\cal A}_{i,\vec k}(t) {\cal A}_{i,-\vec
k}(t')\rangle$ in which the brackets stand for averages in the equilibrium
density matrix.  
A straightforward but tedious exercise following the steps described
in\cite{baymkad,fetter}  
reveals that this relation is given by
\begin{equation}
<<\tilde{{\cal A}}_{Ti}(\vec k, \omega)\tilde{{\cal A}}_{Ti}(-\vec k, -\omega)>>
= \frac{1}{2} \left\{\langle {A}_{i}(\vec k,\omega) {A}_{i}(-\vec k,
-\omega)\rangle  +  \langle {A}_{i}(-\vec k , -\omega)
{A}_{i}(\vec k,\omega)\rangle  
\right\} \label{classicorr} 
\end{equation} 
which again is a result known to be a consequence of the
fluctuation-dissipation 
relation in 
simple systems. The correlation function (\ref{classicorr}) has a finite,
non-trivial classical limit and agrees with the one proposed to be studied 
within the context of classical field theory in \cite{classi7}. 

%new paragraph%%

The real-time analysis presented here agrees with the general
picture discussed in refs.\cite{yaffe,yaff,son}. Moreover, our analysis 
reveals the precise range of the kernels.

Clearly the situation will be more complicated in QCD where a separation between hard and soft degrees
of freedom must be implemented in order to obtain the influence functional for the soft degrees of freedom.
However, the procedure detailed in this section can be carried out consistently once this separation is
introduced.

%end new paragraph%%

Of course the main rationale for obtaining a Langevin equation is to provide a
semiclassical scheme to implement the calculation of correlation functions from
the solutions of stochastic differential equations. However, we note that
unless a successful scheme to deal with the non-Markovian kernels is
implemented the advantages of a Langevin description are at best
formal. Ignoring the non-localities of the dissipative and noise kernels will
clearly miss the important physics associated with Landau damping. A naive
Markovian approximation is not only uncontrolled and unwarranted but clearly
very untrustworthy in view of the fact that the relevant kernels are all
long-ranged and it is precisely this long-ranged nature of these kernels which
is responsible for the important dissipative effects of Landau damping. The
importance and difficulties of keeping these non-localities in a classical
lattice description has been recognized in \cite{humuller,mooremuller,iancu}.

\section{Kinetics of Landau damping} 

The real-time formulation of non-equilibrium quantum field theory allows us to
obtain the corresponding kinetic equations for the relaxation of the 
occupation number or population of quanta. In particular our goal is
to obtain the kinetic equation for the relaxation of the expectation value of
the number of soft
quanta. In keeping with  the focus of this article we
will only consider the population relaxation to lowest order in the HTL
approximation and concentrate mainly on the understanding of relaxation via
Landau damping.

Kinetic approaches towards describing transport phenomena and 
relaxational dynamics typically require  a wide separation between microscopic 
time and length scales, namely the thermal (or Compton) wavelength (mean 
separation of particles) and the relaxation scales (mean free path and 
relaxation time). This approach which ultimately leads to the Boltzmann 
transport equations involves the identification of slow and fast 
variables which justifies a gradient expansion. 
This is a coarse graining procedure that averages over 
microscopic time scales and leads to irreversible time evolution. 
 
In the collisional approach to Boltzmann kinetics only the 
distribution-changing processes that conserve energy and momentum are
considered, and these are weighted by the corresponding  Bose/Fermi statistical
factors. Off-shell processes that occur on time scales  
shorter than the relaxation scale are not included; therefore 
 only processes with asymptotically on-shell final states are accounted for
in this description.
 Landau damping processes which  
contribute via thermal loops to forward scattering  are not included in the
typical Boltzmann equation. 
Thus we anticipate that the naive Boltzmann approach will yield no population
relaxation via Landau damping.  
 However
this is conceptually inconsistent because we have learned in the 
previous sections that an initial coherent configuration will relax to an
amplitude smaller than the initial by these processes and we would expect 
such a relaxation to contribute to a depletion of the number of quanta
of the initial state. The resolution of this inconsistency requires one to go
{\bf beyond} a simple Boltzmann approach and to include off-shell
processes in the kinetic description. This is the focus of this section.  

 The contribution of {\em off-shell} processes to the 
{\em non-equilibrium} evolution of particle distributions is ignored in 
most approaches towards kinetics. 
Recently an approach to kinetics that incorporates off-shell effects has
been proposed to describe processes in which relaxation competes with
other fast scales \cite{photop,boyakinetic}, in particular near phase
transitions \cite{photop}. The importance of off-shell contributions to the
evolution of particle distributions has also been recognized within the context
of fast kinetics in semiconductors \cite{kinetics}.

This section is devoted to a study of kinetics as an {\em initial value 
problem}\cite{photop,boyakinetic,kinetics} in order to reveal the
role played by off-shell processes 
in transport phenomena and relaxation in the medium. 

We {\em derive} a kinetic equation that takes into account  
microscopic time scales in the theory from first principles allowing us
to analyze clearly   
the effect of {\em off-shell} non-collisional Landau damping 
processes on the evolution of the photon distribution. This framework will
allow us to make contact  with the relaxation of a coherent
initial configuration studied in the previous sections. We want to study both the
relaxation of an initial distribution of asymptotic photons with free dispersion relation $\Omega(k)=k$, as
well as for the quasiparticles with dispersion relation $\Omega(k)=\omega_p(k)$ with
$\omega_p(k)$ being the solution of the dispersion equation
(\ref{poletrans}) i.e. the 
`true', in-medium pole.
We will distinguish between these two physically different cases and address
them separately. 

%%%%new paragraph%%%%
This approach begins by  defining a suitable number operator\cite{photop}. In the case of asymptotic photons
this is the usual number operator in terms of the canonical field and momenta which is given by the energy per momentum k
divided by the frequency. In the case of  quasiparticles the energy stored in the plasma has two components: the
free field part plus the response from the medium. From classical electromagnetism of polarizable media\cite{landau} in linear response,  the two contributions lead to an energy density in the medium that is quadratic in terms of the
electric and magnetic fields, each term however, multiplied by a coefficient that involves derivatives of the dielectric
and permeability tensors with respect to frequency\cite{landau}. The contribution from the plasma collective modes is
obtained by evaluating these coefficients at the plasma frequency. 
Using Kramers Kronig dispersion relations\cite{landau}, these coefficients  are related to the residue of the
dielectric constant at the plasma poles. This relation has been formalized at the field theoretical level by
Migdal  in his
pioneering work on collective modes in medium where he developed the
quantization procedure in medium in terms of quasiparticle
operators\cite{migdal}. This field theoretical treatment automatically leads to
the identification of
these coefficients with the residues of the retarded propagators
at the plasmon poles, i.e. the wave function renormalization. Migdal obtains in this manner the energy density corresponding to on-shell collective modes in terms of the quasiparticle operators\cite{migdal}. 

 More recently  the energy density of the plasma including the polarization effects has been obtained in terms of operators that create and destroy collective modes in the plasma by Blaizot and Iancu\cite{blaian}. The
results of these authors is consistent with the collective mode quantization and the energy density obtained by Migdal\cite{migdal} and with the results of classical polarizable media\cite{landau}. Blaizot and Iancu\cite{blaian}
use the collective mode decomposition of the transverse gauge field
\begin{equation}
\vec{A}^t_{\vec k}(t)= \sqrt{\frac{Z^t_k}{2\omega_p(k)}}\sum_{\lambda=1,2}\left[\vec{\epsilon}_{\lambda}(\vec k)a_{\lambda}(\vec k)e^{-i\omega_p(k)t}+\vec{\epsilon}_{\lambda}(-\vec k)a^{\dagger}_{\lambda}(-\vec k)e^{i\omega_p(k)t}\right]
\label{decomcoll}
\end{equation} 
Where the operators $a^{\dagger}_{\lambda}(\vec k)\; ; \; a_{\lambda}(\vec k)$ have a retarded propagator with
unit residue at the plasmon pole and $Z^t_k$ is the wave function renormalization. 

Migdal\cite{migdal} and Blaizot and Iancu\cite{blaian} prove that the energy density associated with the collective modes
in the medium can be written as
\begin{equation}
{\cal E}(k) = \omega_p(k) \sum_{\lambda=1,2}a^{\dagger}_{\lambda}(\vec k)a_{\lambda}(\vec k)
\end{equation}
This result is the same as that obtained in the classical theory of polarizable media when the electric and magnetic
fields (transverse) are written in terms of collective modes\cite{landau}. 

 Thus following Migdal\cite{migdal} and Blaizot and Iancu\cite{blaian} we introduce the Heisenberg
number operator of on-shell collective modes
 
\begin{eqnarray} 
\hat{N}_{\vec{k}}=\frac{1}{4\Omega_k {\cal Z}^t_k}\left[\dot{\vec{A}}_{T}(\vec{k}) 
\cdot\dot{\vec{A}}_{T}(-\vec{k}) 
+\Omega^2_k\vec{A}_{T}(\vec{k})\cdot\vec{A}_{T}(-\vec{k})\right]-\frac{1}{2{\cal Z}^t_k}\;
.  \label{numbergeneral}
\end{eqnarray}
\noindent where for asymptotic photons $\Omega_k =k \; ; \;  {\cal Z}^t_k=1$ (we neglect here the zero temperature contribution) and for collective modes
$\Omega_k =\omega_p(k) \; ; \;  {\cal Z}^t_k= Z^t[T]$ with $\omega_p(k)$ being the plasmon pole
and $ Z^t[T]$ the wave function renormalization in the HTL limit. This formulation now permits to treat
the collective modes much in the same manner as the usual renormalized in and out fields in S-matrix theory, i.e. by
rewriting the action in terms of the renormalized fields and introducing counterterms that reflect the true position
of the pole and  residue. 

%%%%%%%%%%%%%%%end of new paragraph %%%%%%%%%%%%%%%%%%%%%%%%%%%%%%

It turns out to be easier to work with the Heisenberg operator
$\dot{\hat{N}}_k$  
rather than the number operator and is obtained using the Heisenberg operator
equations which are easily seen  
to be:    
\begin{eqnarray} 
&&\dot{\vec{A}}_T(\vec{k},t)=\vec{\Pi}_T(-\vec{k},t)\nonumber \\ 
&&\dot{\vec{\Pi}}_T(-\vec{k},t)=-(k^2+2e^2 
\langle\Phi^\dagger\Phi\rangle+\delta \Omega^2_k)\vec{A}_T(\vec{k},t) 
-\vec{j}_T(\vec{k},t) \; . \nonumber
\end{eqnarray} 
where the counterterm accounts for the definition of the number of
quasiparticles and $\vec{\Pi_T}(\vec k)$ represents the canonical momentum
conjugate to the transverse electromagnetic field  
$\vec{A_T}(\vec k)$. 

%%%new paragraph

We now consider an initial state described by a density matrix for which the {\em expectation value}
of the above number operator for on-shell collective modes is non-vanishing.  

%%%end of new paragraph

Using these equations the expectation value of the Heisenberg rate operator
in the initial density matrix is obtained in 
the following form which is rather convenient for subsequent calculations: 
\begin{eqnarray} 
\langle\dot{\hat{N}}_k\rangle(t)=&&\dot{N}_k(t)=-\frac{1}{2\Omega_k {\cal Z}^t_k} 
\frac{\partial}{\partial t^{\prime\prime}} 
\left[\langle\vec{j}^{+}_T(\vec{k},t)\cdot{\vec{A}^{-}}_T 
(-\vec{k},t^{\prime\prime})  
\rangle\right]_{t=t^{\prime\prime}} \nonumber \\ 
-&&\frac{2e^2\langle\Phi^\dagger\Phi\rangle+\delta \Omega^2_k}{2\Omega_k{\cal Z}^t_k} \; 
\frac{\partial}{\partial t^{\prime\prime}} 
\left[\langle\vec{A}^{+}_T(\vec{k},t)\cdot{\vec{A}^{-}}_T 
(-\vec{k},t^{\prime\prime})+ 
\langle\vec{A}^{+}_T(\vec{k},t^{\prime\prime})\cdot{\vec{A}^{-}}_T 
(-\vec{k},t)   
\rangle\right]_{t=t^{\prime\prime}} \label{ndot2} 
\end{eqnarray} 
This formulation has been previously applied to the study of photon production in a 
strongly out of equilibrium phase transition\cite{photop}.

The expectation values are calculated by inserting the operators into the
closed time path integral and expanding in powers of $\alpha$. Since 
$\delta \Omega^2_k$ is  of order $\alpha$ the second term in
(\ref{ndot2}) is calculated as a tadpole in free field theory and it vanishes
identically. Let us consider the case in which the initial density matrix at
time $t_0$ is diagonal in the basis of eigenstates of the number 
operator, and evolves subsequently with the interaction Hamiltonian. To lowest
order in $\alpha$ we find the expectation value of the rate to be
\cite{photop},  
\begin{eqnarray} 
&&\dot{N_k}(t)= 
\frac{e^2}{4\pi^3\Omega_k {\cal Z}^t_k}\int^t dt^{\prime}\int 
dp\; [p^2-(\vec{p}\cdot{\hat{k}})^2]  
\times
%\label{formalrate} 
\\ \nonumber \\ \nonumber 
&&\times\left 
[G_p^>(t,t^{\prime})G_{k+p}^>(t,t^{\prime})\dot{\cal{G}}^<_k 
(t^{\prime},t)-G_p^<(t,t^{\prime})G_{k+p}^<(t,t^{\prime}) 
\dot{\cal{G}}^>_k(t^{\prime},t)\right]\Theta(t-t^{\prime})\; . \nonumber 
\end{eqnarray} 
The theta function ensures that this expression is causal. The Green's 
functions for the scalars and the photons can be read off from 
Eqs.(\ref{greater}),(\ref{lesser}),(\ref{phot>}) and
(\ref{phot<}) but with the frequency $\Omega_k$ replacing the bare
frequency. Finally the rate can be written as\cite{photop}:  
\begin{eqnarray} 
\dot{N}_{\vec{k}}(t) 
&& =\frac{e^2}{16\pi^3\Omega_k {\cal Z}^t_k}\int \frac{d^3p}{\omega_p\omega_{k+p}}p^2 
\sin^2\theta \int^t_{t_0} d\tau \left\{ \right. \nonumber \\ 
&&\left. \cos[(\omega_p+\omega_{k+p}+\Omega_k)(t-\tau)] 
 \left[(1+N_k(t_0))(1+n_p)(1+n_{k+p})-N_k(t_0) n_p n_{k+p}\right]+ \right.
\nonumber \\
&&\left. \cos[(\omega_p+\omega_{k+p}-\Omega_k)(t-\tau)]
\left[(1+N_k(t_0))n_p n_{k+p}-N_k(t_0) (1+n_p)(1+n_{k+p})\right]+
\right. \nonumber \\ 
&&\left. \cos[(\omega_p-\omega_{k+p}+\Omega_k)(t-\tau)]
\left[(1+N_k(t_0))(1+n_p) n_{k+p}-N_k(t_0) n_p(1+n_{k+p})\right]+
\right. \nonumber \\ 
&&\left. \cos[(\omega_p-\omega_{k+p}-\Omega_k)(t-\tau)]
\left[(1+N_k(t_0))(1+n_{k+p}) n_p-N_k(t_0) n_{k+p}(1+n_p)\right] \right\}
\label{kineticeqn} 
\end{eqnarray}

We note that the expression above depends on the occupation number of the gauge
field at the initial time $t_0$ only -- this is obviously a consequence of
the fact that perturbation theory at lowest order, neglects the change in
occupation number. Recently we have proposed\cite{photop} a Dyson-like
resummation of the 
perturbative expansion that includes off-shell effects in the relaxation of the
population. This resummation scheme is obtained by the replacement $N_k(t_0)
\rightarrow N_k(\tau)$ in (\ref{kineticeqn}) resulting in a non-Markovian
description. The resulting kinetic equation with memory is akin to that
obtained via the generalized Kadanoff-Baym approximation\cite{kinetics} in
non-relativistic many body systems. 
This approximation has been recently shown 
to coincide with the exact result in the weak coupling limit in a solvable
model of relaxation \cite{saeed,nextkinetics} and will be shown below to imply a
Dyson-resummation of the perturbative series. The kinetic
Eq. (\ref{kineticeqn}) has an obvious interpretation in terms of gain minus
loss processes\cite{photop}, but the retarded time integrals and the cosine 
functions replace the more familiar energy conserving delta functions. 
Taking the occupation number outside the integral and integrating to large
times, thereby replacing the cosines by delta functions as in a Boltzmann
description would lead to a vanishing right hand side since none of the resulting energy
conserving delta functions can be satisfied. However, the non-Markovian
kinetic equation (\ref{kineticeqn}) will lead to non-trivial relaxational
dynamics that will be studied in detail below.    

Let us consider the situation in which the initial state has been prepared
far in the past, i.e. $t_0 \rightarrow -\infty$. 

An equilibrium solution is simply 
\begin{equation} 
N_k^{eq}= \text{constant}\;  
\end{equation} 
where the constant is arbitrary  because none of the
resulting energy-conserving delta functions can be satisfied for the values of
$\Omega_k$ either corresponding to the  
bare frequencies or the quasiparticle poles. This is a consequence of the
off-shell processes, a detailed understanding of this feature will be
provided elsewhere\cite{saeed,nextkinetics}. 

Let us now consider departures from this equilibrium solution and study the
relaxation of a disturbance in the distribution function introduced in the
system at $t=0$ so that  
\begin{equation} 
N_k(t=0)=N_k^{eq}+\delta N_k(0).  
\end{equation} 

 Denoting the particle distribution for $t>0$ by 
\begin{equation} 
N_k(t>0)=N_k^{eq}+\delta N_k(t) 
\end{equation}  
we obtain a rate equation for $\delta N_k(t)$ which is  
now the same as in the previous step except that the time integrals stretch
from $0$ to $t$ instead of $t_0 \rightarrow -\infty$ to $t$
\begin{eqnarray} 
&&\frac{d}{dt}\delta N_{\vec{k}}(t)
=\frac{e^2}{16\pi^3\Omega_k {\cal Z}^t_k}\int \frac{d^3p}{\omega_p\omega_{k+p}}p^2 
\sin^2\theta\int^t_{0}d\tau \left\{(1+n_p+n_{p+k}) \right. \nonumber \\
&&\left. \left[\cos[(\omega_p+\omega_{k+p}+\Omega_k)(t-\tau)]-
\cos[(\omega_p+\omega_{k+p}-\Omega_k)(t-\tau)]\right] 
\right.\nonumber \\ 
&&\left. (n_{p+k}-n_p) 
\left[ \cos[(\omega_p-\omega_{k+p}+\Omega_k)(t-\tau)] -
\cos[(\omega_p-\omega_{k+p}-\Omega_k)(t-\tau)] \right] \right\}
\delta N_k(\tau)\; . \label{rateeqn}
\end{eqnarray} 
 
Terms independent of $N_k$ vanish identically since the time integrals
for those terms yield delta functions which are never satisfied.  

Although this non-Markovian but linear equation will be solved exactly by
Laplace transform below, it is illuminating to compare the different
approximations that are obtained under the assumption that the relaxation time
scale for 
$\delta N_k$ is much longer than the time scale of the non-local kernels.
Under this assumption, which will be analyzed below,  $\delta N_k(\tau)$ can be
replaced by $\delta N_k(t)$ and taken outside of the integral leading to a
Markovian description. A further approximation, taking the upper limit of the
remaining integral to $t\rightarrow \infty$ leads to the familiar Boltzmann
equation, thus the 
two approximations to be compared with the `exact' solution are the
following
\begin{itemize}
\item {{\bf Markovian:}
\begin{eqnarray}
&&\frac{d}{dt}\delta N_k^M(t)=   -\Gamma(t) \; \delta N_k^M(t) \label{markovian} \\
&&\Gamma(t) =  - \frac{e^2}{16\pi^3\Omega_k {\cal Z}^t_k}\int
\frac{d^3p}{\omega_p\omega_{k+p}}p^2  
\sin^2\theta\int^t_{0}d\tau \left\{(1+n_p+n_{p+k}) \right. \nonumber \\
&&\left. \left[\cos[(\omega_p+\omega_{k+p}+\Omega_k)(t-\tau)]-
\cos[(\omega_p+\omega_{k+p}-\Omega_k)(t-\tau)]\right] 
\right.\nonumber \\ 
&&\left. (n_{p+k}-n_p) 
\left[ \cos[(\omega_p-\omega_{k+p}+\Omega_k)(t-\tau)] -
\cos[(\omega_p-\omega_{k+p}-\Omega_k)(t-\tau)] \right] \right\}\label{nonmarko}
\; ,
\end{eqnarray} 
with solution 
\begin{equation}
\delta N_k^M(t) = e^{-\int^t_0 \Gamma(t') dt'} \; \delta N_k^M(0)
\label{solumarko} \; .
\end{equation}
} 

\item{{\bf Boltzmann:} 
\begin{eqnarray}
\frac{d}{dt}\delta N_k^B(t) & = &    -\Gamma(\infty) \; \delta N_k^B(t)
\label{boltzmann} \\ 
\delta N_k^B(t) & = &  e^{-\Gamma(\infty)t} \; \delta N_k^B(0)\; .
\label{soluboltz}  
\end{eqnarray}

}
\end{itemize} 

Taking the limit $t\rightarrow \infty$ in $\Gamma(t)$, the cosines become
energy conserving delta functions and comparing with the expression for the  
transverse self energy given by Eq.(\ref{specrepre}), it is straightforward
to see that in the Boltzmann approximation we obtain 
\begin{equation}
\frac{d}{dt}\delta N_k^B(t) \buildrel{t \to \infty}\over=
-\frac{\Sigma_{I}^{t}(\Omega_k)}{\Omega_k} \delta N_k^B(t) \equiv 0\; ,
\label{boltzrate}  
\end{equation}
with $ \Sigma^{t}_{I}(\Omega_k) $ being the imaginary part of the transverse
self-energy evaluated at $\Omega_k$. This result is the familiar relationship
between the  relaxation rate of the particle distribution  $\Gamma(\infty)$ and
the damping rate, which is determined by the imaginary part of the self-energy
{\em on-shell} and vanishes in the present case because the damping 
processes are {\em off-shell}. 

The solution of the non-Markovian equation (\ref{rateeqn}) is obtained
by Laplace transform and given in general by 

\begin{equation}
\delta N_k(t)  =  \delta N_k(0) \int_{c-i\infty}^{c+i\infty} \frac{ds}{2\pi i}
\; \frac{e^{st}}{ s-{\cal S}_k(s) } \label{exactnumb} \; ,
\end{equation}
where $c$ is a real constant chosen so that the contour is to the right of the
singularities of the integrand and ${\cal S}_k(s)$ is the 
Laplace transform of the non-local kernel given by
\begin{eqnarray}
{\cal S}_k(s) & = & \frac{e^2}{16\pi^3 \Omega_k}\int
\frac{d^3p}{\omega_p\omega_{k+p}}p^2 
\sin^2\theta \left\{ \left[ \frac{s}{s^2+\left(\omega_p+\omega_{k+p}+\Omega_k
\right)^2} \right. \right. \nonumber \\ 
&-& \left. \left. \frac{s}{s^2+\left(\omega_p+\omega_{k+p}-\Omega_k
\right)^2}\right](1+n_p+n_{k+p}) \right. \nonumber \\ 
 & + & \left. \left[ \frac{s}{s^2+\left(\omega_p-\omega_{k+p}+\Omega_k
\right)^2}  - \frac{s}{s^2+\left(\omega_p-\omega_{k+p}-\Omega_k
\right)^2}\right](n_{p+k}-n_p) \right\} 
\label{laplanumber}
\end{eqnarray}

The expression (\ref{exactnumb}) clearly shows that the
non-Markovian rate equation (\ref{rateeqn}) implies a Dyson-like resummation of
the perturbative series as anticipated before. 

Comparing the non-Markovian rate equation (\ref{rateeqn}) to the Markovian
approximation (\ref{markovian}) one can clearly see that the Markovian
approximation averages over the time scales of the kernel, whereas the
non-Markovian equation includes the contribution of  coherent processes
throughout the history of the kernel. 
If the range of the kernel was indeed shorter than the relaxation time of the
population, the real-time solutions of both  
equations will differ by terms of the order of the ratio of the
time scale of the kernel to the  relaxation
time scale of the distribution. However, in situations in which the kernel
is long-ranged as is the case under consideration, the non-Markovian expression
allows the inclusion of coherent effects in the relaxation.  

The dominant contribution to ${\cal S}(s)$ in the HTL limit arises from
the Landau damping term leading to the simplified expression

\begin{equation}
{\cal S}_k(s) = -\frac{e^2T^2 \; k\; s}{12\Omega_k} \int^1_{-1}
\frac{x(1-x^2)}{s^2+(kx-\Omega_k)^2} \; dx \label{Shtl} 
\end{equation} 
More explicitly,
\begin{eqnarray}
{\cal S}_k(s) &=&  -\frac{e^2T^2 }{12 k}\left\{ 4 s + {s(k^2 + s^2 -
3\Omega_k^2 )\over 2 k^2 \Omega_k }  \log{s^2 + ( k -\Omega_k)^2
\over   s^2 + ( k +\Omega_k)^2} \right. \cr \cr
&+& \left.
 { k^2 -\Omega_k^2 + 3 s^2 \over 2 i k^2}\log{ (s+ik)^2 + \Omega_k^2
\over  (s-ik)^2 + \Omega_k^2}\right\} \; .
\end{eqnarray}

It is easy to show that this form of   ${\cal S}_k(s)$ is in
fact related to the self-energy in the HTL limit by the following illuminating
equation:
\begin{equation}
{\cal S}_k(s)=\frac{\Sigma^t(s-i\Omega_k)-\Sigma^t(s+i\Omega_k)}{2i\Omega_k}.
\label{kinselfen}
\end{equation}

The real-time solutions to Eqs. (\ref{rateeqn}), (\ref{exactnumb}) and
(\ref{laplanumber}) in the HTL
limit with (\ref{Shtl}) will now be given for the cases of quasiparticle and
bare particle respectively. This analysis will 
reveal the range of validity of the assumptions leading to the Markovian
approximation. 

\subsection{Quasiparticle: $\Omega_k=\omega_p(k)$} 

When $\Omega_k$ is chosen to be the quasi-particle pole, $\omega_p(k)>k$ we
note that the integrand  
of (\ref{exactnumb}) has a single isolated pole at $s=0$. Indeed, the limit of
$s\rightarrow 0$ in ${\cal S}_k(s)$ would lead to delta 
functions in Eq.(\ref{Shtl}) which, however, cannot be satisfied for
$\Omega_k=\omega_p(k)$.  Therefore, $ {\cal S}_k(0) = 0 $ and
$ s=0 $ is an isolated single pole and completely determines the
asymptotic limit of the real-time solution. This can also be seen  
by looking at Eq. (\ref{kinselfen}), from which the analytic structure is
explicit. Clearly, ${\cal S}_k(s)$ vanishes at $s=0$ because
$\Sigma^t(-i\omega_p(k))=\Sigma^t(i\omega_p(k))$. Furthermore from the known
singularity structure of $\Sigma^t(s)$ one concludes that ${\cal S}_k(s)$ must
have branch cuts for $i(\omega_p(k) -k)<s<
 i(\omega_p(k) +k)$ and $-i(\omega_p(k) -k)<s< -i(\omega_p(k) +k)$.  
Using Eq. (\ref{kinselfen}), it is now a straightforward 
exercise to see that the residue at this pole at $s=0$  is
given by $(1-2\partial\Sigma^t(i\Omega)/\partial\Omega^2)^{-1}$ which to this
order is $\approx Z^t[T]^2$ where $Z^t[T]$ the (transverse) wave function
renormalization given in Eq. (\ref{zetatrans}).

 Again the
 long time behavior is completely dominated by the end points of the cut,
 leading to the asymptotic result 
\begin{eqnarray}  
\delta N_k(t)&\buildrel{t \to \infty}\over =&\delta N_k(0)\left\{ Z^t[T]
+ { e^2T^2 \pi^2 \over {12 Z^t[T] \,  k \, \omega_p(k) \, t^2}} 
\left[ { \cos{(\omega_p(k)+k)t}
\over (\omega_p(k)+k)^2\; \left( 1 + {  {e^2T^2 D_+ }
\over {3 k^2}} \right)} \right.  \right. \cr \cr
&-&\left.\left. { \cos{(\omega_p(k)-k)t}
\over (\omega_p(k)-k)^2\; \left( 1 + {  {e^2T^2 D_-} \over {3 k^2}} \right) }
\right]\right\}
\label{asintoquasipart}
\end{eqnarray}  
%+{\cal O}\left(\frac{e^2T^2}{k^2}\right)
%\frac{\cos[(\omega_p(k)-k)t]}{t^2(\omega_p(k)-k)^2}+\right.
%\\\nonumber
%&&\left.+{\cal O}\left(\frac{e^2T^2}{k^2}\right)
%\frac{\cos[(\omega_p(k)+k)t]}{t^2(\omega_p(k)+k)^2}\right] 

where
$$
D_{\pm} \equiv 1 + \left( \frac12 \pm { \omega_p(k) \over k} \right)
\log\left( 1 \pm { k  \over {  \omega_p(k)}}  \right) \; .
$$
We clearly see that asymptotically the population has relaxed to a smaller
value and the ratio of the asymptotic to the initial population is
determined by the square of the thermal wave-function renormalization. 
This is in agreement with the analysis of the relaxation of the 
expectation value of the field -- since $N_k \propto A_T^2$ it is expected that
the ratio of the asymptotic value of the quasi-particle population to the initial
value be proportional to the square of the same relation for the expectation
value of the field. 
%%%new paragraph

Therefore the relaxation of the quasiparticle number has its origin in Landau damping, this is consistent
with the results of Blaizot and Iancu\cite{blaian} who proved that the time derivative of the total
energy is related to Landau damping. Since the quasiparticle number is related to the energy of the collective
modes the relaxation of the quasiparticle number is directly related to Landau damping. 

%%%end new paragraph

This intuition is also borne out in the usual Boltzmann
approach wherein the relaxation rate of the distribution function (in the
relaxation time approximation) is twice the damping rate for the
quasiparticle. 

%%%%new paragraph
This relaxation has a simple interpretation.  Consider the case of a physical electron with interpolating
operators defined to create single electrons with the physical mass and unit amplitude out of the  in or out
vacuum states. The asymptotic correlation function in real time of these interpolating operators has the
oscillatory parts corresponding to the physical pole (with unit residue), but there are power law corrections arising from the overlap of the states created by these operators with the multiparticle continuum\cite{brown}. Whereas at zero temperature
the multiparticle continuum is beyond the two particle threshold, at finite temperature and in the case under consideration, the leading contribution is
obtained from Landau damping corresponding to intermediate states with space-like momenta. 
%%end of new paragraph

For the Markovian approximation (\ref{solumarko}) we find for 
$ t\rightarrow \infty$
\begin{equation}
-\int^t_0 \Gamma(t')dt' \approx 2 {\frac{\partial \Sigma_t(\omega)}{\partial
\omega^2}}|_{\omega_p(k)}+ {\cal O}\left[ 
\left(\frac{e^2T^2}{k^2}\right)\left(
\frac{\cos[(\omega_k-k)t]}{t^2(\omega_k-k)^2}+\frac
{\cos[(\omega_k+k)t]}{t^2(\omega_k+k)^2}\right)\right] 
\label{markosolu}
\end{equation}
For $e^2T^2/k^2<<1$ we see that to lowest order in $e^2T^2/k^2$ the
perturbative expansion of the Markovian solution  coincides with the solution
of the non-Markovian equation. However for soft momenta such an expansion
is not valid and the validity of the Markovian approximation must be 
questioned.   

\subsection{Failure of the Markovian and Boltzmann approximation:}
To assess whether the Markovian and Boltzmann approximations will be reliable we
must understand the different time scales, in particular the range of 
the kernel. 

In the Hard Thermal approximation we find that the non-Markovian kinetic
equation (\ref{rateeqn}) reduces to,  
\begin{eqnarray} 
\frac{d}{dt}\delta N_k(t)\simeq-\frac{e^2T^2k}{12\Omega_k}\int_{-1}^{1}dx\;(1-x^2)x 
\int_0^td\tau\; \sin[kx(t-\tau)]\sin[\Omega_k(t-\tau)]\; \delta N_k(\tau). 
\end{eqnarray} 
The integral over the variable $x$ inside the kernel can be performed and
we find that the kernel falls off as $1/(t-\tau)^2+\cdots$. Then if $eT/k <<1$
the relaxation time scale of the population is longer than the range of the
kernel and the Markovian approximation is warranted. In this case the
discrepancies between the non-Markovian and Markovian results are
perturbatively small. On the other hand, for soft scales the relaxation 
time scales become comparable to the time scale of the kernel and a Markovian
approximation is certainly unjustified. The non-Markovian equation for
relaxation includes the coherent effects on similar time scales that 
are averaged out (coarse-grained) in the Markovian approximation.

\subsection{Bare particle, or hard quasiparticle:}

For the case of the bare particle, the dispersion relation is simply
$\Omega_k=k$ -- this is also the case for the large $k$ limit of the 
quasiparticle dispersion relation\cite{htlnonabel,weldon1,lebellac}. In
this 
case, the quantity ${\cal S}_k (s)/s$ in (\ref{laplanumber}) has a logarithmic
singularity as $s\rightarrow 0$, because the position of the putative 
pole $\Omega_k$ has moved to the tip of the cut and there is a pinching
singularity. There is no longer a pole at $s=0$ in the Laplace transform 
$\left[s-{\cal S}_k(s)\right]^{-1}$, rather it diverges as $(s\ln s)^{-1}$ as
$s\rightarrow 0$. This logarithmic divergence arising from the pinching 
singularity is very similar to that recently studied within the context of
hard fermions\cite{infra}. 

In the Markovian approximation (\ref{markovian}) and in the Hard Thermal Loop
limit, the rate equation becomes 
\begin{eqnarray} 
&&\frac{d}{dt}\delta N_{\vec{k}}(t) 
=\frac{e^2}{4\pi^2}\int dp \frac{dn_p}{dp}\int_{-1}^{1}dx 
(1-x^2)x p^2\int^t_{0}d\tau\left 
\{\cos[(k(1-x)(t-\tau)]\right\}\delta N_k(t)\nonumber \\ 
&&\simeq-\frac{e^2T^2}{12k}\left[\frac{\sin(2kt)}{k^2t^2} 
+\frac{2}{kt}-\frac{2}{k^3t^3}+\frac{2\cos(2kt)}{k^3t^3}\right]\delta
N_k(t)\label{kinehtlbare}.  
\end{eqnarray}  
which for long times yields a power law with an anomalous exponent: 
\begin{equation} 
\delta N_k(t)\sim \delta N_k(0)(kt)^{-\frac{e^2T^2}{6k^2}}.\label{anomexp} 
\end{equation} 
An anomalous exponent somewhat similar to this one has been found
in\cite{infra} in the case of a hard fermion and it has the same 
origin, i.e. a pinching infrared singularity. The expression
(\ref{kinehtlbare}) reveals that the kernel is long ranged, falling off 
with an inverse power of time in this case. Therefore a Markovian approximation
will be justified  when $e^2T^2/k^2 <<1$ because only in 
this weak coupling limit is the population relaxation {\em slower} than the
fall off of the kernel.

The solution of the non-Markovian equation (\ref{rateeqn}, \ref{laplanumber})
in the HTL limit (only the Landau damping contribution is considered) is
obtained again by inversion of the Laplace 
transform. We find that the long time behaviour is given by the end-points. In
this case the $\omega = 0 $ end-point dominates yielding 
\begin{eqnarray} 
\delta N_k(t)&=&\delta N_k(0)\; \frac{e^2T^2}{6\,k^2} Re \int_0^{\infty}
{{dy}\over y} {{e^{-y}}\over{\left[ 1 +
\frac{e^2T^2}{6\,k^2}\log{{2kt}\over {iy {\bar e}}} \right]^2 + \left( {{\pi
e^2T^2}\over { 12\,k^2}}\right)^2}}\cr \cr
&&\left[ 1+{\cal O}\left(\frac{1}{t}\right)\right] 
% &\buildrel{t \to \infty}\over= &\frac{e^2T^2}{12\,k^2}{1 \over
%{\frac{e^2T^2}{3\,k^2}\log^2 
%kt}}\left[1+{\cal O}\left(\frac{1}{\log t}\right)\right]\; .
\end{eqnarray} 
where $ {\bar e} = 2.718\ldots $ is the base of the natural
logarithms.

It is illuminating to point out that as $\Omega_k \rightarrow k$ the
logarithmic singularities in the real part of the transverse self energy 
imply that (for fixed finite $k$)
\begin{equation}
\frac{\partial \Sigma^t(\omega)}{\partial \omega^2}|_{\omega \rightarrow k}
\rightarrow \infty \; \; \Longrightarrow \; \;  
Z^t[T] \rightarrow 0 \label{zerowave}
\end{equation}

As the pole approaches the tip of the cut (for finite $k$) the residue becomes
smaller until it vanishes exactly when the pole merges with the continuum.  
It is remarkable that in this limit distortions of the distribution function
relax completely and vanish asymptotically. Although for the case of a hard
quasiparticle with $k >> eT$ the dispersion relation approaches that of  a free
particle,  $Z^t[T] \rightarrow 1$ and physics is perturbative, there is,
however, slow relaxation.   
We note that none of these
effects can be captured by a simple Boltzmann approach since all of these
phenomena are associated with off-shell effects. 

In the case of the distribution function for bare particles, we can
interpret this anomalous relaxation as the dressing effect from the 
medium, i.e. at long times the bare particles are completely dressed
by the medium and disappear from the spectrum. 

The interpretation is different for the case of a hard quasiparticle,
in which case the free dispersion relation is obtained in the limit
$k >> eT$. In this limit the effective coupling $eT/k <<1$ and the relaxation
is slow. 
%%%%%%%%new paragraph%%%%%%%%
This result could be important in understanding the relaxation of a distribution of photons produced
in Bremmstrahlung processes at high energy in the quark gluon plasma. 
%%%%%%%%end of new paragraph%%%%%%%%%%%
A full  study of the relaxation of hard quasiparticles is beyond
the realm of this article and will be studied in detail
elsewhere\cite{nextkinetics}.

\section{Discussion and Conclusions}

Our goal in this article was to provide a detailed analysis of the real-time
relaxation of soft gauge invariant non-equilibrium expectation value through the off-shell process of
Landau damping. These 
determine the leading contributions to the thermal propagators in the Hard
Thermal Loop limit and are the dominant contributions to the long-time
asymptotics. The off-shell nature of these processes determine the
non-Markovian nature of the relaxation phenomena associated with them. 
We focussed our study on the leading HTL contributions to the relaxation of
gauge invariant transverse and longitudinal non-equilibrium expectation value in scalar
electrodynamics. These results will also apply to fermion electrodynamics and
non-abelian theories since the structure of the retarded Green's function is
the same up to this order in the HTL expansion. After providing 
an elegant  re-derivation of the well known HTL effective action using
the tools  
of non-equilibrium quantum field theory in the linear amplitude 
approximation\cite{inhomodcc}, we moved on to the main goals of this
article:

%%ADDITION 9 HERE

\begin{itemize}
\item{ {\bf i:}
to study in detail the relaxation in real time, in the
linear approximation (small amplitude) of the transverse and longitudinal
gauge invariant non-equilibrium expectation value. The off-shell process of Landau damping
results in power law relaxation of the transverse and logarithmic relaxation of
the longitudinal (plasmon or charge-density) excitations. Both types of 
non-equilibrium expectation value relax to an asymptotic amplitude that depends on the thermal
wave-function renormalization, which is completely determined by Landau damping
in the HTL limit. One of the main conclusions of this detailed
analysis is that the relaxational dynamics asymptotically at long
times and to leading order in  HTL resummation is completely
determined by the behavior of the spectral density near {\em the
Landau damping thresholds} 
at $\omega = \pm k$ (a branch cut singularity), the contribution from
the region $\omega \approx 0$ are regular (no branch point
singularities) and therefore lead to subleading corrections to
dynamics at long times and high temperature. This result is for both
longitudinal and transverse 
non-equilibrium expectation value and is confirmed by an exhaustive analytic
and numerical study. The short time evolution of the non-equilibrium expectation value is
determined by moments of the total spectral density. Therefore a
complete understanding of the 
the global analytic structure of the retarded propagator, in particular the
complete cut contribution from Landau damping processes is required. 

 This is special
to the HTL resummation at one-loop: at higher orders,  a
branch point could develop at $\omega = 0$.  Such
a branch point would produce pure power like tails, with no oscillations,
and so dominate at large times.

We restricted ourselves in this paper to small amplitude
non-equilibrium expectation value so that we were confined to the linear regime. New phenomena beyond
the HTL  
scheme are expected in the non-linear  amplitude regime. Such
regimes can be studied within  self-consistent
Hartree-type approximations in the out of equilibrium framework\cite{disip}. }

\item{ {\bf ii:} 

We have obtained the influence functional (non-equilibrium effective action)
for the soft gauge invariant degrees of freedom by integrating out the hard
degrees of freedom to leading order in the HTL approximation. This allowed us
to obtain the Langevin equation for the soft
degrees of freedom to leading order in HTL  
and to provide a microscopic {\em ab initio} calculation of the dissipative and
noise kernels in the HTL limit. Both kernels display 
the non-localities associated with Landau damping and we find that 
there is no region of time scales in which a Markovian
approximation describes the dynamics correctly. As a byproduct we obtained
the fluctuation-dissipation relation and recognized the correlation function
that emerges in the classical limit. We established in detail  that a
Markovian description of relaxation of transverse 
or longitudinal non-equilibrium expectation value is unwarranted to this order in the HTL
resummation. } 

\item{{\bf iii:} Having understood the relaxation of coherent non-equilibrium expectation value
through off-shell effects of Landau damping we asked how these processes
can be incoporated in the relaxation of the distribution function for
the transverse fields. A Boltzmann approach would yield no relaxation to
lowest order in the HTL because there is no imaginary part on-shell for
the transverse or longitudinal quasiparticles. Therefore a kinetic description of the relaxation of the distribution function must necessarily go beyond a Boltzmann collision approximation. 

We provided a novel description of the kinetics of relaxation of
the distribution function for transverse degrees of freedom that includes
off-shell effects and goes far beyond the usual Boltzmann approach. We have
compared the relaxation obtained from this non-Markovian kinetic equation to
that described by the usual Boltzmann (which yields a trivial result) and a
Markovian version that includes coarse-grained details of the off-shell 
processes. We have found that the distribution function for soft
quasiparticles relaxes as a power $(1/t^2)$, and found an unusual dressing
dynamics for bare particles. This kinetic approach also 
reveals unusual logarithmic real time relaxation for hard quasiparticles
resulting from infrared pinching singularities similar to those found in
the case of hard fermions\cite{infra}.}

\end{itemize}

The body of these results reveals new and unusual features of relaxation
of soft degrees of freedom in gauge theories. These will obviously have
to be taken seriously into account in a full description of relaxational
processes in the QGP and should also be important to clarify better the
role played by damping in the sphaleron rate in the symmetric phase. 

In higher order corrections there will arise contributions from collisional
processes that provide
an imaginary part on-shell. At next order for
example both Compton scattering and pair-annihilation will contribute
to collisional relaxation and will provide a collisional width both to the
transverse and longitudinal (plasmon) degrees of freedom. The imaginary
part of the self-energy on-shell is typically associated with a damping
rate and associated with an exponential decay of the amplitude. Excepting
intermediate time scales, this exponential relaxation, however is not a proper
description either at 
early or long times where power laws dominate the dynamics
\cite{inhomodcc}. The relaxation associated with Landau damping at 
lowest order, will have to be balanced with the next order corrections which
yield 
an approximate exponential relaxation and the resulting dynamics will depend on
the details of the competition between these different processes, both 
Landau damping and collisional. Which process dominates will depend on the
particular time scale of interest and the time scales for competition between
the two different type of phenomena will depend on the details of the
perturbative contributions. The kinetic approach introduced here could also
prove 
useful to study the energy loss of quarks and leptons via off-shell processes
in the QGP.   

We plan to address this competition between Landau damping and collisional
phenomena, along with an extension of the treatment presented in this article
to leptons and a more detailed study of non-Markovian kinetics in future work.

\section{acknowledgements}
D. B. thanks the N.S.F. for partial support through grant PHY-9605186, he also thanks J. Smit,
G. Aarts, P. Arnold, D. Son, W. Buchmuller, A. Jakovac and A. Patkos for conversations.
R. H. and S. P. K. were supported by DOE grant DE-FG02-91-ER40682.
S. P. K. would like to thank BNL for hospitality during the progress of this work. 
The authors acknowledge support from NATO. 

%%%%%%%%%%%%%%%%%%%%% BEGIN REFERENCES %%%%%%%%%%%%%%%%%%%%

%%%%%%%%%%%%%%%%%% END REFERENCES %%%%%%%%%%%%%%%%%%%%%%%%%%%

%%%%%%%%%%%%%%%% BEGIN FIGURES %%%%%%%%%%%%%%%%%%%%%%%%%%%%%%%

%%%%%%%%%%%%%%END FIGURES %%%%%%%%%%%%%%%%%%%%%%%%%%%%%%%%%%%%%%%

\end{document}